\documentclass[notitlepage,amsmath,amssymb,aps,reprint,superscriptaddress]{revtex4-1} 
\usepackage{graphicx}
\usepackage{dcolumn}
\usepackage{bm}
\usepackage{braket}
\usepackage{amsmath}
\usepackage{bm}
\usepackage{amssymb}
\usepackage{setspace}
\usepackage[euler]{textgreek}
\allowdisplaybreaks
\usepackage{multirow}
\usepackage{comment}
\usepackage{relsize}
\usepackage{hyperref}
\usepackage{siunitx}
\hypersetup{
    colorlinks=true,
    linkcolor=blue,
    filecolor=magenta,      
    urlcolor=cyan,
    citecolor=red
}
\usepackage[x11names]{xcolor}
\usepackage{appendix}

\newcommand{\treset}{\tau_{\text{reset}}}
\newcommand{\ron}{r_{\text{on}}}
\newcommand{\roff}{r_{\text{off}}}
\newcommand{\ssin}{s_{\text{in}}}
\newcommand{\sout}{s_{\text{out}}}
\newcommand{\neff}{n_{\text{eff}}}
\newcommand{\adag}{a^{\dagger}}
\newcommand{\bdag}{b^{\dagger}}
\newcommand{\bsdag}{B^{\dagger}}

\begin{document}

\title{Scalable and High-Fidelity Quantum Random Access Memory\\ in Spin-Photon Networks }

\author{K. C. Chen$^{1,2}$, W. Dai$^{1,3}$, C. Errando-Herranz$^1$, S. Lloyd$^{1,4}$, D. Englund$^{1,2}$\\
$^1$\textit{Research Laboratory of Electronics, M.I.T., Cambridge, Massachusetts 02139, USA}\\
$^2$\textit{Department of Electrical Engineering and Computer Science, M.I.T., Cambridge, Massachusetts 02139, USA}\\
$^3$\textit{Department of Computer Science, U. Mass., Amherst, Massachusetts 01003, USA}\\
$^4$\textit{Department of Mechanical Engineering, M.I.T., Cambridge, Massachusetts 02139, USA}}
\begin{abstract}
A quantum random access memory (qRAM) is considered an essential computing unit to enable polynomial speedups in quantum information processing. Proposed implementations include using neutral atoms and superconducting circuits to construct a binary tree, but these systems still require demonstrations of the elementary components. Here, we propose a photonic integrated circuit (PIC) architecture integrated with solid-state memories as a viable platform for constructing a qRAM. We also present an alternative scheme based on quantum teleportation and extend it to the context of quantum networks. Both implementations rely on already demonstrated components: electro-optic modulators, a Mach-Zehnder interferometer (MZI) network, and nanocavities coupled to artificial atoms for spin-based memory writing and retrieval. Our approaches furthermore benefit from built-in error-detection based on photon heralding. Detailed theoretical analysis of the qRAM efficiency and query fidelity shows that our proposal presents viable near-term designs for a general qRAM.
\end{abstract}
\maketitle

\section{Introduction}
\begin{figure}[h!]
    \centering
    \includegraphics[width=0.5\textwidth]{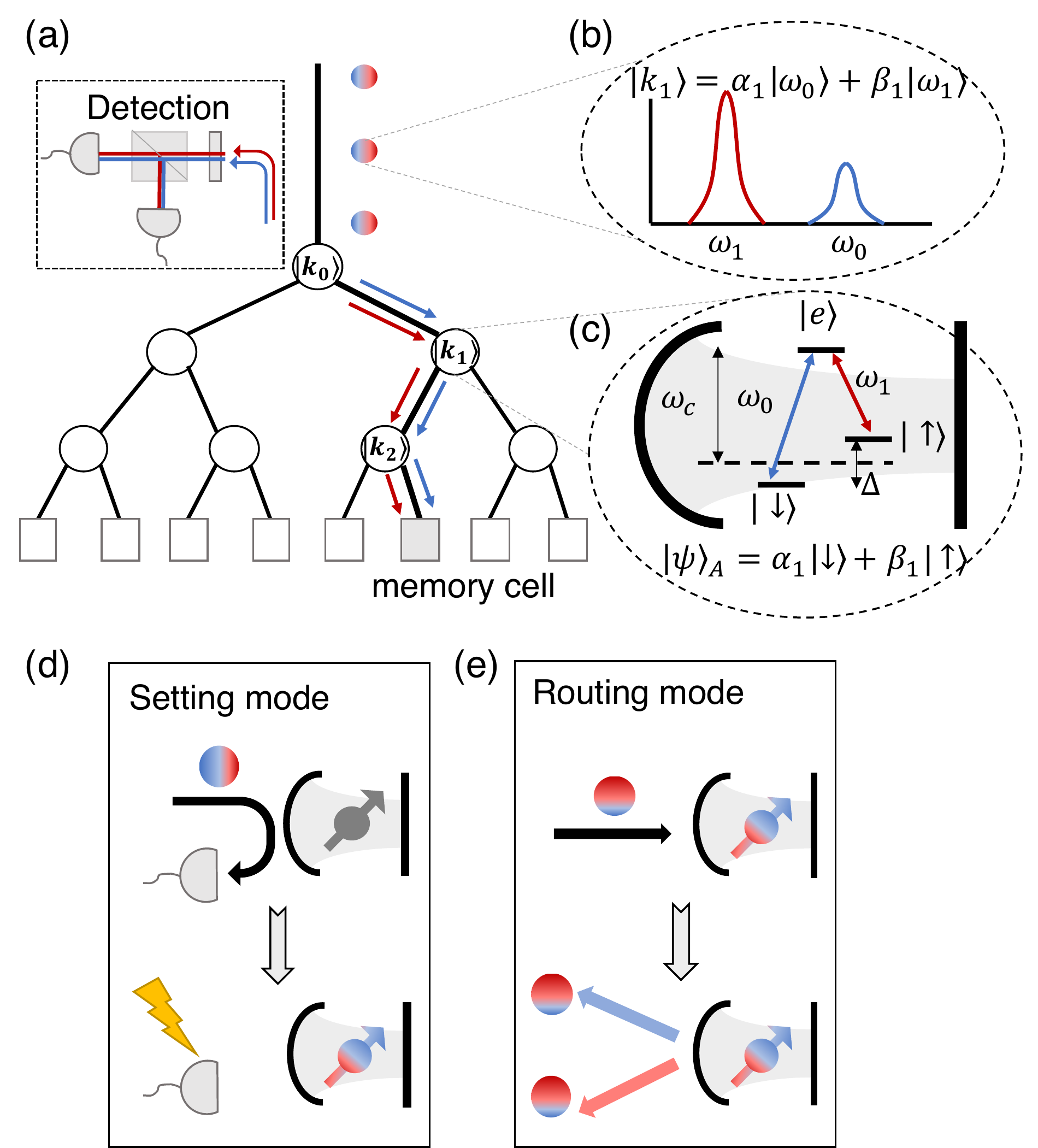}
    \caption{An illustrative bucket-brigade model with a cavity-coupled $\Lambda$-level atom at each tree node. (a) The address $\ket{j}$ consisting the register qubits $\ket{k_0}\ket{k_1}\ket{k_2}$ arrives at the $3$-level binary tree containing $N=2^3$ memory cells. (b) Each register is a frequency-encoded photonic qubit in the $\{\omega_0,\omega_1\}$ basis. (c) For our implementation, each tree node is a $\Lambda$-atom coupled to a single-sided nanocavity whose resonant frequency $\omega_c$ is tuned to the average of the two atomic transition frequencies, $\omega_0$ and $\omega_1$, which are separated by a Zeeman splitting $\Delta$. For layer 1, the register $\ket{k_1}$ sets the node's internal state to $\ket{\psi_A}=\alpha_1\ket{\downarrow}+\beta_1\ket{\uparrow}$ that routes the successive register $\ket{k_2}$. Two essential operations are (d) the setting mode via cavity reflection and (e) the routing mode.}
    \label{fig:setup}
\end{figure}

Random access memory (RAM) is a fundamental computing unit that allows on-demand storing and retrieving data. While a classical RAM addresses one memory cell in the database per operation, a quantum RAM permits querying a superposition of multiple memories ~\cite{Giovannetti_2008_PRL}. Given a superposition of addresses $j$, the `qRAM' returns a correlated set of data $D_j$:
\begin{align}
\ket{\psi}_{\text{in}}=\sum_{j=1}^N \alpha_j \ket{j}_a\ket{\emptyset}_b \xrightarrow{\text{qRAM}} \ket{\psi}_{\text{out}}=\sum_{j=1}^N \alpha_j \ket{j}_a\ket{D_j}_b
\end{align}
where $N$ is the number of memory cells and the subscripts $a$ and $b$ denote the address and bus qubits, respectively. One efficient implementation of qRAM proposed by Giovannetti, Lloyd, and Maccone (GLM) ~\cite{Giovannetti_2008_PRL,Giovannetti_2008_PRA} is the `bucket-brigade model': a binary tree of memory nodes that direct the bus qubit to the data layer. Each preceding layer $i$ in a tree of depth $n$ represents the register $k_i$ of the address $\ket{j}=\ket{k_1k_2...k_{n-1}k_n}$, which sets the path leading to the corresponding memory cell $D_j$ (Fig.~\ref{fig:setup}(a)).

Principally, these memory nodes must (1) store an address register qubit that (2) routes ensuing qubits for addressing and retrieval. The register $\ket{k_i}$ sets layer $i$'s internal state that governs routing of the subsequent registers $\{\ket{k_{i+1}},\ket{k_{i+2}},...\}$. A qRAM query thus performs a sequence of alternating state transfer and routing operations, with each register qubit determining how the node routes the subsequent register. Once the binary tree has been programmed by the state of address qubits, $\sum_j\alpha_j\ket{j}_a$, it is traversed by the bus photon $\ket{\downarrow}_b$ to access the memory cells $\{D_j\}$ in superposition. The bus qubit travels back up the tree and addresses are mapped onto the returning register qubits to disentangle themselves from the nodes, producing the qRAM output state $\ket{\psi}_{\text{out}}$. The ability to perform this operation in log($N$) time steps highlights the advantage of quantum parallelism and offers polynomial speedups in quantum algorithms for applications such as quantum machine learning~\cite{Biamonte_2017}, matrix inversion~\cite{Harrow_2009}, quantum imaging~\cite{Kiani_2020}, and quantum searching~\cite{Grover_1996}.

Despite its mathematical elegance, proposed implementations of qRAM have not been experimentally demonstrated. The existent proposals are based on neutral atoms~\cite{Giovannetti_2008_PRA,Hong_2012,Moiseev_2016} and superconducting circuits~\cite{Hann_2019}, but still requiring elementary components to be realized. Here, we introduce a scheme that assembles separately demonstrated technologies into a photonic integrated circuit (PIC) architecture integrated with artificial atoms. Namely, the system contains a high-fidelity frequency beam splitter~\cite{Lu_2018,Joshi_2020,Lu_2020}, nanocavities strongly coupled to long-lived spin memories~\cite{Nguyen_2019_PRB,Bhaskar_2019}, and a scalable nanophotonic Mach-Zehnder interferometer (MZI) array~\cite{Harris_2017}. Importantly, the protocol relies on a cavity-assisted controlled-phase (CZ) gate~\cite{Duan_2004} whose heralding inherently provides the ability to detect qubit loss. We estimate our PIC implementation of the GLM scheme to achieve efficiency of kHz query rate for a qRAM containing $>10^2$ memory cells. Furthermore, we propose an alternative approach based on quantum teleportation. This teleportation scheme enables scaling to $10^5$ memories and still achieving $>$kHz query rate. More importantly, the protocol's framework applies to quantum networks that require no additional modifications. Thus, our study provides a promising blueprint for building a general qRAM essential for quantum information processing.

\section{Architecture}
In our PIC implementation, the address register and the bus qubits are frequency-encoded photons $\ket{\psi_P}=\alpha\ket{\omega_0}+\beta\ket{\omega_1}$ prepared by a frequency beam splitter~\cite{Lu_2018,Joshi_2020,Lu_2020} shown in Fig.~\ref{fig:setup}(b). They arrive at each node in the binary tree and interact with a cavity-coupled atom, which has two spin states $\ket{\downarrow}$ and $\ket{\uparrow}$. Both states are coupled to an excited state $\ket{e}$ with respective transition frequencies $\omega_0$ and $\omega_1$ shown in Fig.~\ref{fig:setup}(c). In this proposal, we specifically consider diamond's negatively charged silicon-vacancy (SiV$^-$) center strongly coupled to a single-sided cavity~\cite{Nguyen_2019_PRB,Bhaskar_2019}. By having the cavity resonance $\omega_c$ equally detuned from the two transitions, i.e. $\omega_{0,1}=\omega_c\pm\Delta/2$ where $\Delta$ is the Zeeman splitting between the spin states, the resulting Fano interference satisfies the following conditions: upon a cavity reflection, the photon acquires no phase shift when it is resonant with the atomic transition; otherwise, it receives a $\pi$ phase shift (see App.~\ref{sec:atom_cavity}).

\begin{figure}[h!]
    \centering
    \includegraphics[width=0.5\textwidth]{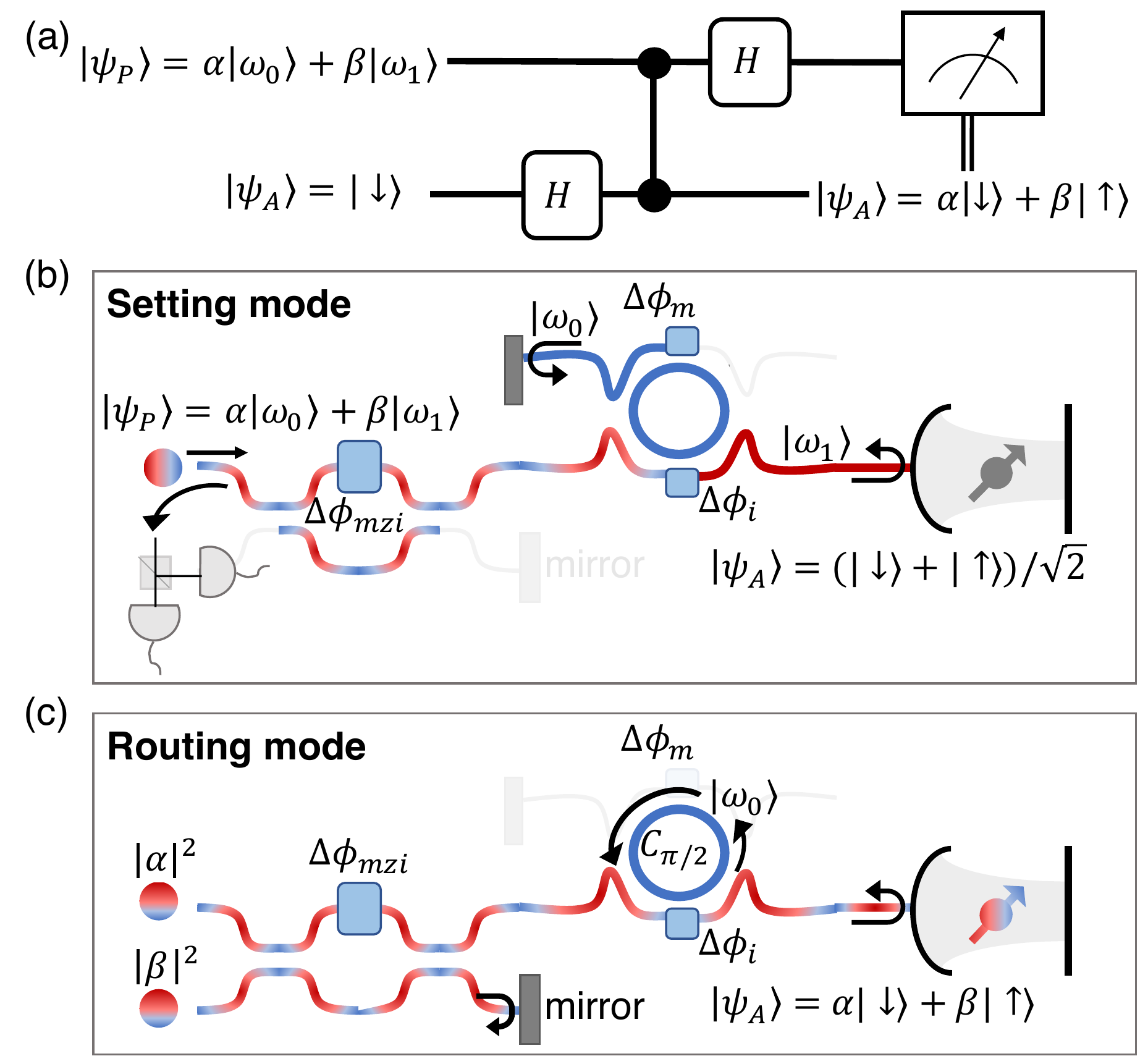}
    \caption{PIC implementation of qRAM. (a) The circuit representation of a quantum state transfer operation that maps the register qubit $\ket{\psi_P}$ onto the atomic qubit $\ket{\psi_A}$. (b) In the setting mode, the photon undergoes a CZ operation to complete quantum state transfer. After passing through the MZI, the $\ket{\omega_0}$ component resonantly couples to the add-drop filter that imparts a $\pi$ phase shift upon reflection off the mirror, while the $\ket{\omega_1}$ component interacts with the atom-cavity system and acquires a spin-dependent phase shift. (c) In the routing mode, the MZI is set to a 50:50 beam splitter, and the top waveguide of the add-drop filter is decoupled such that the ring resonator imparts a $\pi/2$ phase shift to the $\ket{\omega_0}$ component upon a single pass. After cavity reflection, the returning photon re-interferes with itself and is routed to either the $\ket{\downarrow}$ path with probability $|\alpha|^2$ or the $\ket{\uparrow}$ path with probability $|\beta|^2$.}
    \label{fig:operations}
\end{figure}

This spin-dependent phase shift enables the two operation modes necessitated by the bucket-brigade model: the photonic qubit ``setting'' the spin state (Fig.~\ref{fig:setup}(d)) and the spin qubit routing the subsequent register qubits (Fig.~\ref{fig:setup}(e)). The cavity interaction enables a CZ gate for heralding a quantum state transfer between the photonic and the spin qubits, as shown in Fig.~\ref{fig:operations}(a). The very same phase dependence on the atomic state also allows quantum routing by leveraging the cavity system as an interferometer.

Explicitly in the PIC platform, each node comprises an MZI, an add-drop filter resonant with the $\omega_0$ component, and a single-sided nanocavity coupled to an SiV$^-$ center. First, in the setting mode, the atom is initialized in a superposition state $\ket{\psi_A}=(\ket{\downarrow}+\ket{\uparrow})/\sqrt{2}$ by a Hadamard operation. Fig.~\ref{fig:operations}(b) shows the register qubit $\ket{\psi_P}=\alpha\ket{\omega_0}+\beta\ket{\omega_1}$ arriving at the MZI and exiting out of the top output port. An add-drop filter then directs the $\omega_0$ component to a mirror (e.g. Sagnac loop reflector) such that $\ket{\omega_0}$ acquires a $\pi$ phase shift upon reflection regardless of the spin state. On the other hand, the $\omega_1$ component continues down the path and reflects off the atom-cavity system, acquiring a spin-dependent phase shift. Finally, the $\omega_0$ and $\omega_1$ components recombine and undergo a Hadamard transformation by a frequency beam splitter before heralding the completion of quantum state transfer. It is essential for the detection system to be shared by all the qRAM layers at the root of the tree. A local detection would otherwise unveil the path information and thereby collapse the superposition of addresses. In other words, all the register qubits must reflect off the qRAM nodes and return to the root to preserve entanglement between the spin qubits and the address paths. 

After the photon detection, the MZI is switched to a 50:50 beam splitter and the tunable add-drop filter is turned ``off'' such that the ring resonator only imparts a $\pi/2$ phase shift to the $\omega_0$ component upon a single pass (see App.~\ref{sec:add_drop_filter}). Hence, the photon acquires a spin-dependent phase shift \textit{independent} of the frequency component. Illustrated in Fig.~\ref{fig:operations}(c), the subsequent register qubit $\ket{k_1}$ arrives at the 50:50 beam splitter. One of the MZI output ports connects to the same path as before, while the other leads to a mirror. As a result, the photon taking the former route acquires a spin-dependent phase shift from interacting with the cavity while one taking the latter route always acquires a $\pi$ phase from reflecting off the mirror. The returning photon then interferes with itself at the beam splitter and is routed to an exit port depending on the spin state. With $|\alpha|^2$ probability, the photon exits out of the top path corresponding to the $\ket{\downarrow}$ spin state; and with $|\beta|^2$ probability, it travels down the bottom path corresponding to the $\ket{\uparrow}$ spin state. Effectively, the beam splitter in conjunction with the atom-cavity system constitute an MZI with the spin-cavity system acting as a phase shifter.

Both the setting and routing operations are repeated alternatingly, carving out the path for the bus qubit to arrive at the desired memory cells. The data can be transferred onto the bus qubit with the same cavity reflection scheme by reversing the role of the photonic and the spin qubits, followed by a projective measurement on the atom. Finally, the sequence is run backwards to disentangle the binary tree from the address qubits, leaving the data qubits $\ket{D_j}$ correlated with their respective addresses $\ket{j}$. 

\subsection{Fidelity} \label{sec:fidelity}

In our cavity-assisted scheme, qubit loss is a heralded error. Therefore, a sequence of successful photon detection guarantees the absence of infidelity stemming from photon loss in the qRAM output. Here, we analyze imperfections in the atom-cavity system that critically affects quantum state transfer as the primary sources of infidelity in our protocol, since any inexact mapping from the register qubit to the spin qubit would result in faulty routing of the subsequent registers. To characterize the setting fidelity given an input register $\ket{\psi}_P=\alpha\ket{\omega_0}+\beta\ket{\omega_1}$, we calculate the resultant spin state $\ket{\psi}_A$ after heralding via a Schrodinger picture evolution:
\begin{align}
\ket{\psi}_A &= (2\alpha r_m\pm\beta(\ron+\roff))\ket{\downarrow}\mp\beta(\ron-\roff)\ket{\uparrow}
\end{align}
where $\ron$ ($\roff$) is the on-resonance (off-resonance) cavity reflectivity and $r_m$ is the mirror reflectivity.

Its overlap with the target state $\ket{\psi}_A=\alpha\ket{\downarrow}+\beta\ket{\uparrow}$ defines the state transfer fidelity $\mathcal{F}$, of which we take the average over six representative states $\ket{\phi_i}$ (axes of a Bloch sphere)~\cite{Chuang_2010,Bowdrey_2002}:
\begin{align}
\mathcal{F} &= \frac{1}{6}\sum_i \mathcal{F}_i = \frac{1}{6}\sum_i \left|\langle\phi_i|\psi_{s,f}(i)\rangle\right|^2.
\end{align}
where $\ket{\phi_1}=\ket{\downarrow},\ket{\phi_2}=\ket{\uparrow},\ket{\phi_{3,4}}=(\ket{\downarrow}\pm\ket{\uparrow})/\sqrt{2},\ket{\phi_{5,6}}=(\ket{\downarrow}\pm i\ket{\uparrow})/\sqrt{2}$ in the $\{\ket{\downarrow},\ket{\uparrow}\}$ basis. 

\begin{figure}[h!]
    \centering
    \includegraphics[width=0.5\textwidth]{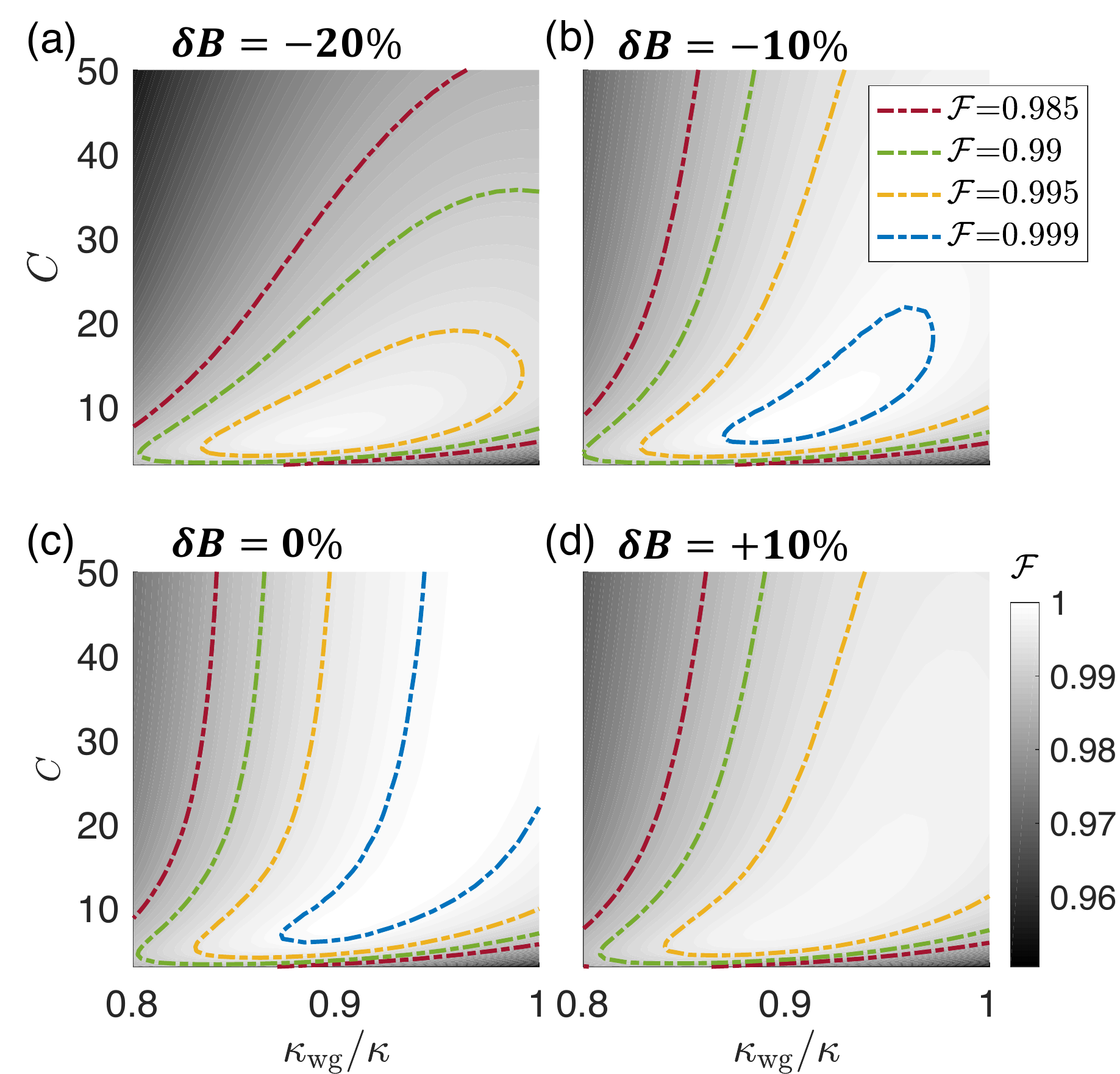}
    \caption{Quantum state transfer and qRAM query fidelities. The transferred state fidelity for a single setting operation is plotted against the atom-cavity cooperativity $C$ and the waveguide-cavity coupling strength $\kappa_{\text{wg}}/\kappa$ for magnetic field deviations (a) $\delta B=-20\%$, (b) $-10\%$, (c) $0\%$, and (d) $10\%$. The contour lines denote the fidelity thresholds at $\mathcal{F}=0.985,0.99,0.995,0.999$.}
    \label{fig:fidelity}
\end{figure}

The setting mode's performance relies on the cavity's coupling strength to the output waveguide mode. When the waveguide-cavity coupling is unity, i.e. $\kappa_{\text{wg}}/\kappa=1$, the cavity reflection solely determines the transfer fidelity that scales as $(C-1)/(C+1)$ in the large cooperativity limit~\cite{Hu_2008,Tiecke_2014}. However, for any reduced $\kappa_{\text{wg}}/\kappa<1$, the need to balance losses becomes especially important. For example, for a desired state $\ket{\phi_3}$ where $\alpha=\beta=1/\sqrt{2}$, balancing losses entails matching the moduli of the on- and off-resonance cavity reflectivities $\ron\propto \kappa_{\text{wg}}(C-1)/(C+1)$ and $\roff\propto \kappa_{\text{wg}}/\kappa$ (see App.~\ref{sec:state_transfer}).

In Fig.~\ref{fig:fidelity}, we analyze $\mathcal{F}$ as a function of $\kappa_{\text{wg}}/\kappa$, $C$, and $\delta B$, which is the deviation from the optimal magnetic field $B_{\text{opt}}\propto \sqrt{\gamma\kappa\left(2C+\kappa\left(\kappa-\kappa_{\text{wg}}\right)/4-\gamma^2/4\right)}$ for the suitable Fano line-shape (see App.~\ref{sec:atom_cavity}). For each point in the fidelity contour, a particular value of $r_m$ is chosen to optimize the fidelity assuming the mirror is tunable. When $\delta B=-20\%$, Fig.~\ref{fig:fidelity}(a) indicates that only a selective range of $C\lessapprox 20$ and $\kappa_{\text{wg}}/\kappa\in\{0.83,0.98\}$ result in $\mathcal{F}>0.995$. However, as the magnetic field deviation reduces to $-10\%$ from the optimum, the transferred state fidelity can exceed $0.999$ for a selected range of $C$ and $\kappa_{\text{wg}}/\kappa$. Fig.~\ref{fig:fidelity}(c) shows that at the optimal magnetic field, i.e. $\delta B=0\%$, the transfer fidelity well exceeds 0.999 for any $C>20$ and $\kappa_{\text{wg}}/\kappa>0.94$. Interestingly, a small region of cooperativities $C<20$ and $\kappa_{\text{wg}}/\kappa<0.94$ can still achieve $\mathcal{F}>0.999$ by carefully balancing losses. However, the tolerance to a varying $C$ diminishes as $\kappa_{\text{wg}}/\kappa$ decreases. As $\delta B$ approaches 10$\%$, however, the setting fidelity can no longer reach 0.999. Its disparity with $\delta B=-10\%$ stems from the asymmetry exhibited by Fano interference (see App.~\ref{sec:atom_cavity}).

\subsection{Efficiency} \label{sec:efficiency}

Next, we analyze the qRAM efficiency by first calculating the success probability of heralding each register qubit $\ket{k_i}$ and then the average rate of completing a single query call. Recall that for the bus qubit to reach the memory layer in an $n$-level qRAM, each register photon $\ket{k_i}$ must travel to the node in layer $i\in\{1,...,n\}$ and return to the detector after cavity reflection. Given a propagation loss $\eta_p$, the probability of completing the round-trip without loss is $e^{-\eta_p L(i)}$, where $L(i)$ is twice the distance between the layer $i$ and the root node. However, since the photon can scatter off the single-sided cavity and the mirror into non-waveguide modes, interaction at each layer further reduces the probability of detecting the returning register qubit by $R_{\text{cav}}$ and $R_m$, which represent the cavity and mirror reflection coefficients, respectively. We take their mean reflection coefficient and define the setting efficiency as $\eta_s=\eta_{\text{det}}(R_m+R_{\text{cav}})/2$, where $\eta_{\text{det}}$ is the detection efficiency. Similarly, the routing efficiency for each layer $i$ would be $\eta_r=R_{\text{cav}}$ assuming lossless transmission through the interferometric coupler. As a result, the probability of successfully heralding each register $\ket{k_i}$ is:
\begin{align}
p_i &= e^{-\eta_p L(i)}\eta_r^{i-1}\eta_s \quad \text{for}\ i\in\{1,...,n\}
\end{align}

\begin{figure}[h!]
    \centering
    \includegraphics[width=0.5\textwidth]{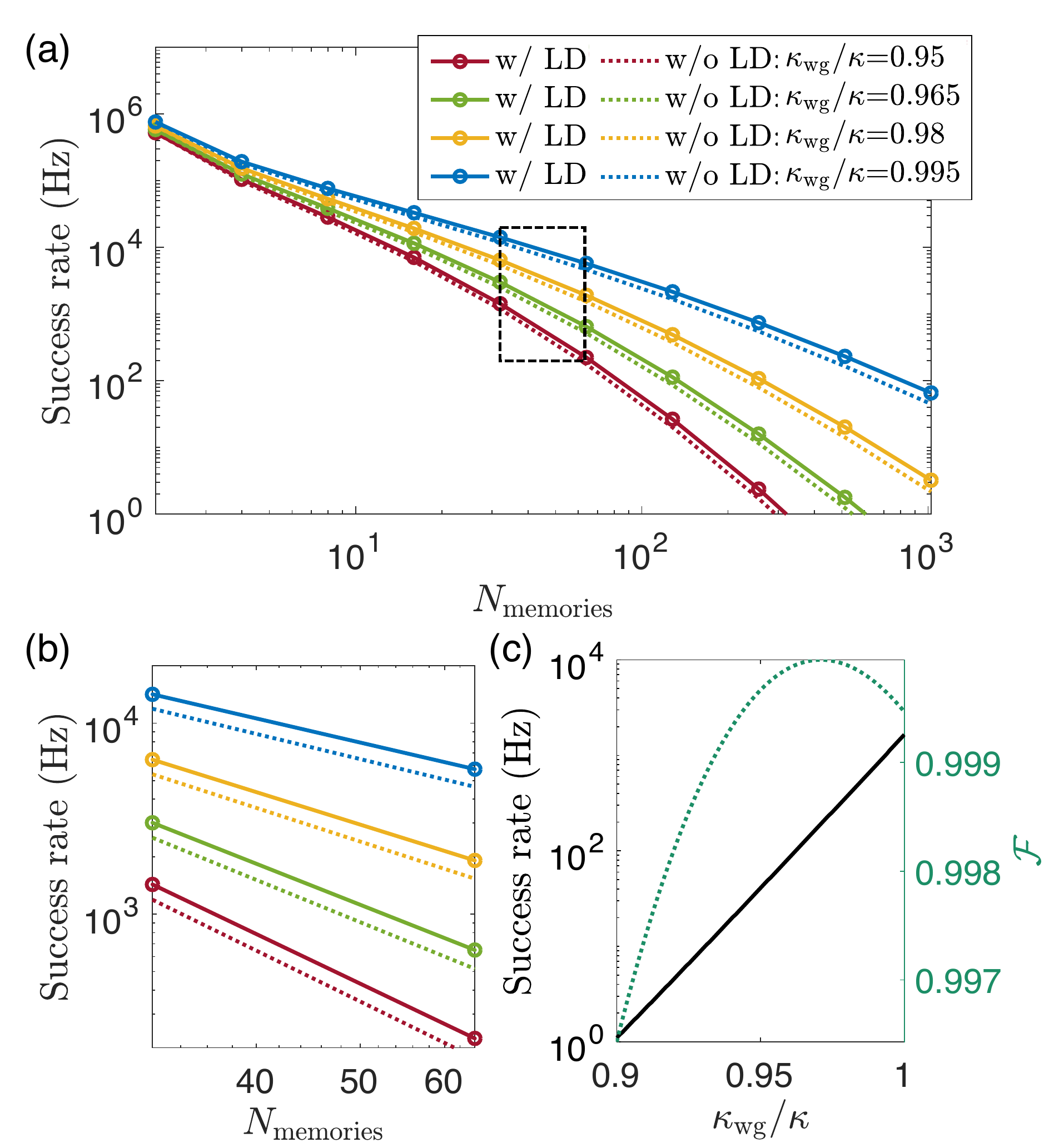}
    \caption{Efficiency of the PIC qRAM. (a) The success rate (Hz) is plotted against $N_{\text{memories}}=2^n$ for a $n$-level qRAM for $\kappa_{\text{wg}}/\kappa=0.95,0.965,0.98,0.995$ for schemes with (solid) and without (dashed) qubit loss detection (LD). On a log-log scale, the success rate rolls off polynomially with increasing $N_{\text{memories}}=2^n$ due to an exponentially decreasing success probability of setting each layer $i$ (see App.~\ref{sec:success_probability}). (b) A zoom-in plot of the black box in (a), highlighting the slight gain in efficiency for the cavity-assisted scheme with LD. (c) Both the success rate and transfer fidelity vary as a function of $\kappa_{\text{wg}}/\kappa$. For a 6-level qRAM with $C=100$, there exists a trade-off between $\bar{\Gamma}$ and $\mathcal{F}$ after $\kappa_{\text{wg}}/\kappa\approx 0.97$ where $\mathcal{F}$ is maximized by perfectly balancing losses.}
    \label{fig:efficiency}
\end{figure}

To calculate the success rate, we must now include both the round-trip travel time to each layer $i$ denoted as $t_i=L_{\text{PIC}}(i)/v_{g,\text{PIC}}+L_{\text{dmd}}(i)/v_{g,\text{dmd}}$, where $L_{\text{PIC}}$ ($L_{\text{dmd}}$) and $v_{g,\text{PIC}}$ ($v_{g,\text{dmd}}$) are the travel distance and group velocity in the PIC (diamond) waveguide. The average time until a successful query call can be found by using the linearity of expectation value. For example, the expected time for a 2-level qRAM is:
\begin{align}
\bar{T}_{n=2} &= p_1p_2(t_1+t_2)+(1-p_1)(\bar{T}_{n=2}+t_1+\treset)\nonumber\\
&+p_1(1-p_2)(\bar{T}_{n=2}+t_1+t_2+\treset)
\end{align}
where $\treset=5~\mu$s is the spin reset time. The first term on the right-hand side is the case of no photons being lost, thus its expected time is simply the product between the success probability of two consecutive heralds $p_1p_2$ and the total travel time $t_1+t_2$. The next term represents the case of the $k_1$ register photon being lost before detection with probability $1-p_1$. Consequently, the average query time $\bar{T}_{n=2}$ is penalized by the additional time $t_1+\treset$. Similarly, if the $k_1$ photon is heralded but the subsequent register $k_2$ is lost with probability $p_1(1-p_2)$, $\bar{T}_{n=2}$ is lengthened by $t_1+t_2+\treset$. Solving for $\bar{T}_{n=2}$ yields:
\begin{align}
\bar{T}_{n=2} &= \frac{t_1+\treset}{p_1p_2}+\frac{t_2}{p_2}-\treset
\end{align}
The expression can be treated as a summation of each layer's round-trip time weighted by its correspond geometric mean, subtracted by $\treset$ since the final trial is a successful run without the need to reset.

We can generalize the average time for a $n$-level qRAM:
\begin{align}
\bar{T} &= \left(\prod_i p_i\right) \left(\sum_i t_i\right)+(1-p_1)(\bar{T}+t_1+\treset)\nonumber\\
&+p_1(1-p_2)(\bar{T}+t_1+t_2+\treset)+...\nonumber\\
&+\left(\prod_i^{n-1}(1-p_n)\right)\left(\bar{T}+\sum_i t_i+\treset\right)\\
\Rightarrow \bar{T} &= \left(\sum_{i=1}^n \frac{t_i}{\prod_{j=i}^n p_j}\right)+\frac{\treset}{\prod_{j=1}^n p_j}-\treset
\end{align}

Finally, the query rate is then:
\begin{align}
\bar{\Gamma} &= \frac{1}{\bar{T}}
\end{align}

Fig.~\ref{fig:efficiency}(a) shows the qRAM query rate as a function of the number of memories $N_{\text{memories}}=2^n$ for different waveguide-cavity coupling $\kappa_{\text{wg}}/\kappa=0.95,0.965,0.98,0.995$. As $N_{\text{memories}}$ increases, the rates roll off polynomially on the log-log scale since the success probability $p_{\text{succ}}$ diminishes super-exponentially with increasing $n$ (see App.~\ref{sec:success_probability}). Furthermore, $p_{\text{succ}}$ intimately depends on the cavity reflection coefficient $R_{\text{cav}}\propto\kappa_{\text{wg}}/\kappa$, causing $\bar{\Gamma}$ to vary drastically with the waveguide-cavity coupling. For example, the difference between $\kappa_{\text{wg}}/\kappa=0.95$ and $\kappa_{\text{wg}}/\kappa=0.995$ exceeds more than an order of magnitude for $N_{\text{memories}}>10^2$, and the disparity grows exponentially as the circuit depth $n$ increases. The unforgiving drop-off in the success rate emphasizes the need for a highly over-coupled single-sided cavity in our protocol. 

On the other hand, our cavity-assisted scheme's built-in loss detection enables a slight boost in success rate. For a scheme without such loss detection, the qRAM must complete the entire sequence of setting and routing all $n$ register qubits \textit{before needing to reset}, assuming qubit loss has occurred and been detected after the query. The corresponding success rate would be:
\begin{align}
\bar{\Gamma}_{\text{no LD}} &= \bar{T}^{-1}_{\text{no LD}} = \left(\frac{\sum_i t_i +\treset}{\prod_i p_i}-\treset\right)^{-1}
\end{align}
In contrast, our protocol periodically checks for register losses via photon detection. Therefore, time can be saved by halting and immediately resetting the spins as soon as quantum state transfer fails to herald. Note that the gain in rate, however, depends on the ratio between travel time $t_i$ and $\treset$. Fig.~\ref{fig:efficiency}(b) shows a modest increase in success rate for our scheme with $t_i<1~\mu$s and $\treset=5~\mu$s relative to one without loss detection. If $\treset\gg t_i$, the slight improvement in efficiency would dwindle as $\bar{\Gamma}$ converges to $\bar{\Gamma}_{\text{no LD}}$.

Lastly, due to the need to balance losses to achieve high transfer fidelity as noted in Sec.~\ref{sec:fidelity}, there exists an inevitable fidelity-rate trade-off. Given a qRAM containing $2^6$ memory cells, Fig.~\ref{fig:efficiency}(c) shows that $\mathcal{F}$ reaches its maximum at $\kappa_{\text{wg}}/\kappa\approx 0.97$. However, the success rate still increases monotonically with $\kappa_{\text{wg}}/\kappa$ even past this optimum fidelity point. The waveguide-cavity coupling regime in which the trade-off exists narrows with higher atom-cavity cooperativity, since both $|\ron|$ and $|\roff|$ increase with $C$ and $\kappa_{\text{wg}}/\kappa$. Nonetheless, at $C=100$ (which has been experimentally demonstrated in Ref.~\cite{Nguyen_2019_PRB,Bhaskar_2019}), the success rate can already exceed $1~$kHz while maintaining high fidelity $\mathcal{F}>0.999$.

\section{Teleportation scheme}

While the aforementioned scheme is viable for a low-depth qRAM, the need to \textit{sequentially} set each address register via cavity reflection inhibits scaling up to $10^6$ memories due to photon loss from cavity interaction. Here, we present an alternative approach that writes the address registers onto all the layers \textit{simultaneously} via quantum teleportation. Crucial to this step is the ability to perform high-fidelity two-qubit gate operation locally between an electron spin (broker qubit) and its neighboring nuclear spin (memory qubit). We assume $>0.99$ gate fidelity to be easily achievable via composite pulses and optimal classical control~\cite{Rong_2015}.

\begin{figure}[h!]
    \centering
    \includegraphics[width=0.5\textwidth]{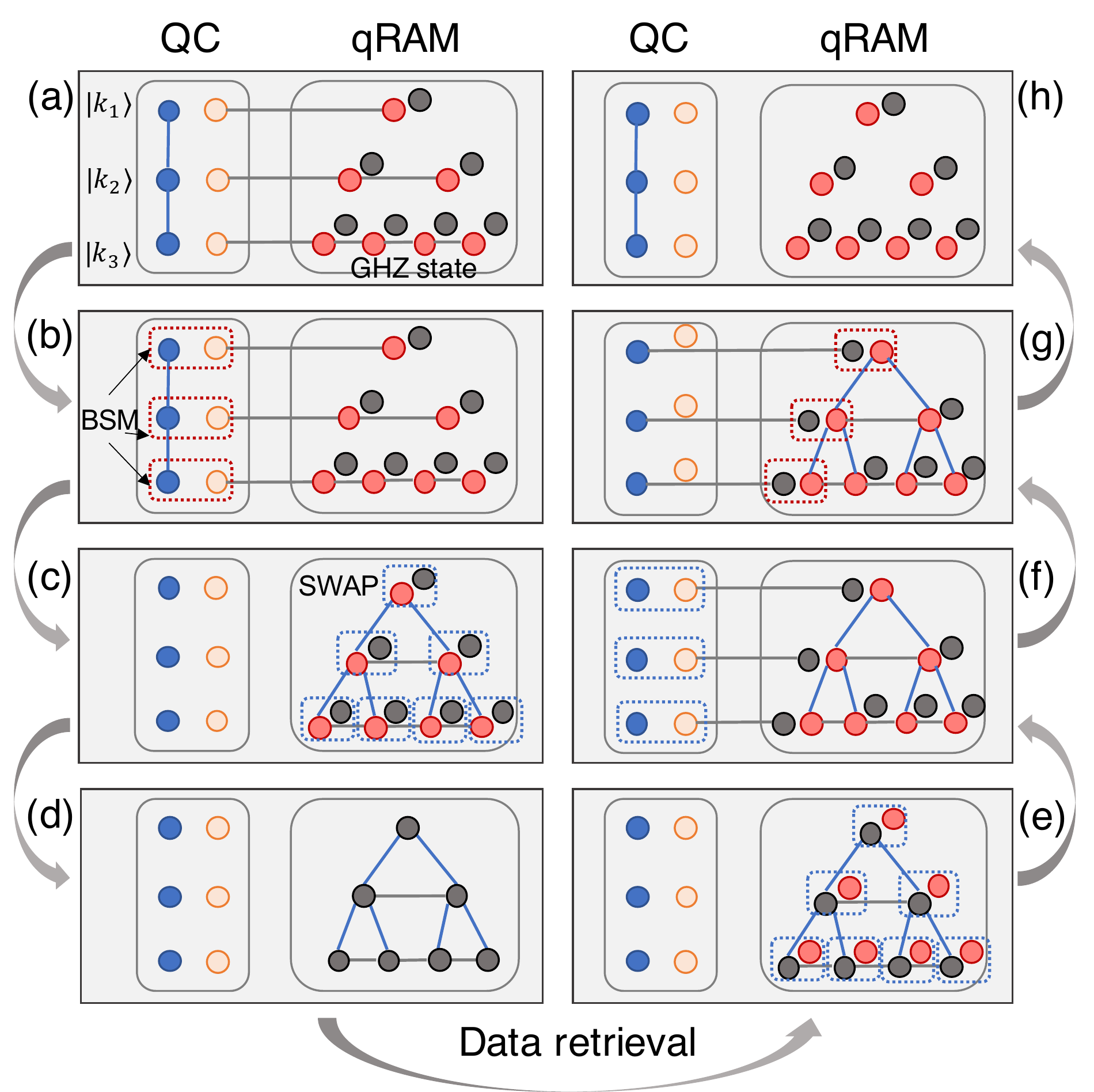}
    \caption{A step-by-step procedure of the teleportation scheme. A quantum computer (QC) holds the query addresses that would be mapped onto a remote qRAM. (a) The QC and qRAM ancillary qubits are remotely entangled (as represented by connecting gray lines), and each qRAM layer's nodes are entangled in a GHZ state. (b) Local bell state measurements (BSM) and subsequent Pauli transformations teleport the query addresses onto the binary tree (c) Then, in each node, the memory (red circle) and the broker (gray circle) qubits undergo a SWAP operation, leaving (d) the qRAM ready for the data retrieval process. (e) After the bus qubit has completed querying, the registers are swapped back onto the memory qubits. (f) After the ancillary qubits are remotely entangled, local SWAP operation prepares entanglement between QC's memory qubits (blue circles) and the qRAM. (g) Local BSM in the qRAM then teleport the query addresses back onto the QC, returning (h) the binary tree to its original state.}
    \label{fig:teleportation_scheme}
\end{figure}

Let us suppose the qRAM is spatially separated from a quantum computer (QC) that holds the list of query addresses $\sum_j \alpha_j\ket{j}_a$. Each register $\ket{k_i}$ lives in a nuclear spin that is a long-lived memory qubit and is accompanied by an electron spin acting as a broker qubit $\ket{\psi_{a,\text{QC}}}$. The electron spin is remotely entangled with another broker qubit $\ket{\psi_{a,\text{qRAM}}}$ in the qRAM via photon-assisted Bell state creation (see App.~\ref{sec:bell_state_creation}). All the nodes (except the leftmost node that is entangled with the QC) \textit{within each layer} $i$ in the binary tree are first initialized as a GHZ state: $\ket{\Psi_{i}}=\left(\ket{00...0}+\ket{11...1}\right)/\sqrt{2}$. Then, a local SWAP operation between the nuclear and electron spins (see App.~\ref{sec:spin_swap}) in each layer's leftmost node produces GHZ states spanning both the QC and the qRAM, as depicted in Fig.~\ref{fig:teleportation_scheme}(a).

Subsequently, each register qubit undergoes a local Bell state measurement (BSM) followed by conditional Pauli transformations, as illustrated in Fig.~\ref{fig:teleportation_scheme}(b). As a result, the query addresses are teleported onto the qRAM for all the layers \textit{simultaneously} (Fig.~\ref{fig:teleportation_scheme}(c)). Prior to sending the bus qubit to the qRAM, all the nodes undergo a SWAP operation between the memory and the broker qubits, resulting in the final entangled network shown in Fig.~\ref{fig:teleportation_scheme}(d).

The data retrieval process remains the same as before. Starting from the root node, the bus photon propagates down the binary tree and is routed based on the state-dependent cavity reflection at each layer. After which, the addresses are swapped onto the memory qubits ((Fig.~\ref{fig:teleportation_scheme}(e)), followed by generating remote entanglement between $\ket{\psi_{a,\text{QC}}}$ and $\ket{\psi_{a,\text{qRAM}}}$ ((Fig.~\ref{fig:teleportation_scheme}(f)). Then, a local SWAP operation entangles the QC's memory qubits with the qRAM. Finally, local BSMs for all the qRAM layers teleport the query addresses back onto the QC ((Fig.~\ref{fig:teleportation_scheme}(g)), returning the binary tree in its waiting state for subsequent queries ((Fig.~\ref{fig:teleportation_scheme}(h)).

Importantly, the proposed architecture extends beyond a PIC platform and can be run on a \textit{quantum network}, in which each network node represents a tree node in the qRAM. Distillation can be used to generate high-fidelity Bell states~\cite{Kalb_2017}, which are then joined to form the GHZ states in the same fashion as heralding entanglement links in a quantum repeater. The protocol's modularity effectively allows the qRAM query to act as a subroutine for distributed quantum computing.

\subsection{Efficiency comparison}

\begin{figure*}
    \centering
    \includegraphics[width=0.8\textwidth]{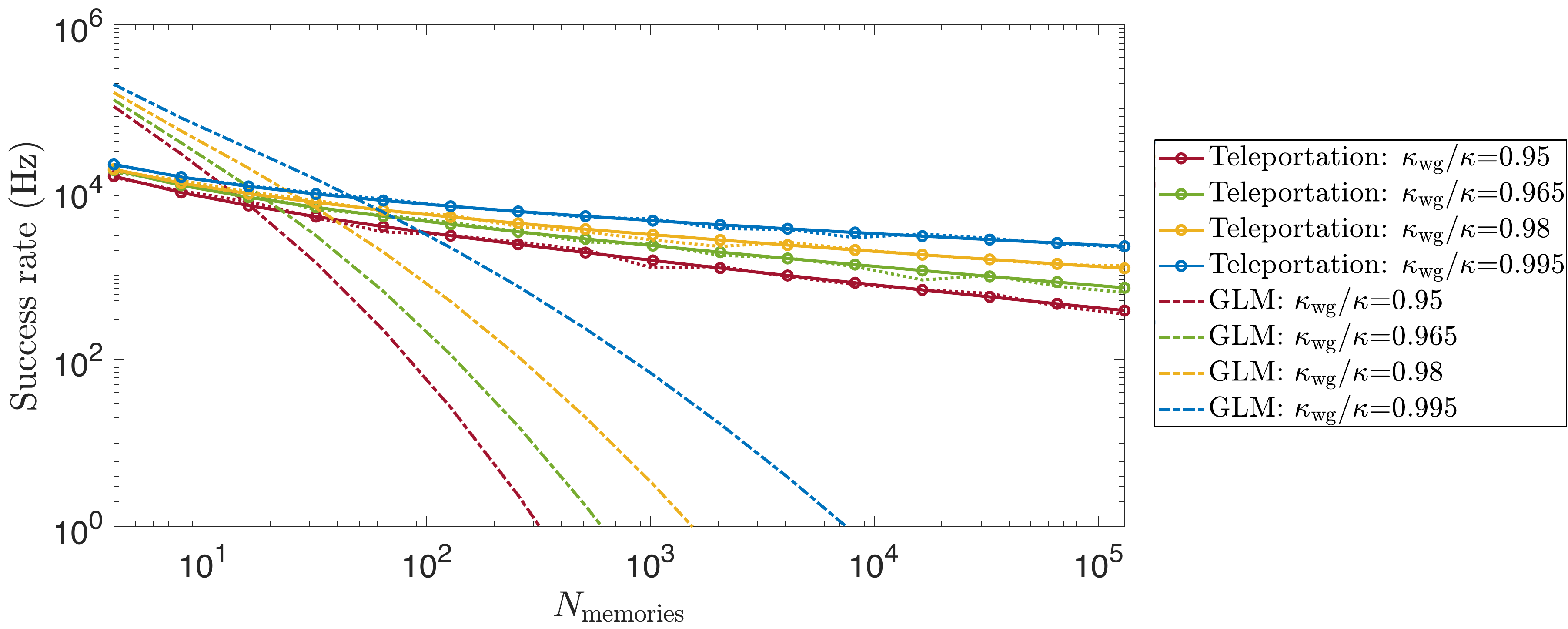}
    \caption{Efficiency comparison between the conventional GLM scheme (dashed dot) and the teleportation scheme. For the teleportation scheme, the solid lines are analytical fits to the simulation data represented by the dashed lines (see App.~\ref{sec:efficiency_simulations}). Each scheme is evaluated at different cavity-waveguide coupling strengths $\kappa_{\text{wg}}/\kappa=0.95,0.965,0.98,0.995$.}
    \label{fig:teleportation_efficiency}
\end{figure*}

Here, we compare the efficiency of the two proposed schemes. The teleportation approach, similar to the GLM scheme, still requires restarting the query procedure if the bus photon is lost during the retrieval step since the path information is revealed by the environment. Despite which, the rate of success for the teleportation scheme still scales much more favorably than the GLM approach. Fig.~\ref{fig:teleportation_efficiency} compares the query efficiency between the two approaches. For small circuit sizes $<10^2$ memories, the GLM scheme achieves higher success rates since the process of generating GHZ states and remote entanglement links is more costly in time than directly transferring the registers sequentially. However, as the qRAM depth increases past the crossover region with $\sim 10^2-10^3$ memories, the GLM scheme's efficiency rolls off rapidly.

On the other hand, the teleportation scheme's success rate decreases relatively slowly. Its efficiency is primarily constrained by the retrieval step that succeeds with probability $\propto\eta_r^n$, as opposed to $\propto\eta_r^{n(n-1)/2}\eta_s^n$ in the GLM scheme. Its favorable scaling is conducive to increasing the circuit size for general-purpose applications such as quantum machine learning~\cite{Biamonte_2017}. Our efficiency simulations (see App.~\ref{sec:efficiency_simulations}) show that the teleportation-based approach can theoretically achieve an average $>$kHz query rate for a qRAM containing $10^5$ memories.

\subsection{Query fidelity}

One potential drawback of the teleportation approach is its requirement to prepare a GHZ state, whose decoherence rate increases linearly with its size. Here we consider the worst case where the entirety of the binary tree is active, meaning all possible addresses are used. Assuming the coherence times of the electron~\cite{Sukachev_2017} and nuclear spins~\cite{Abobeih_2018} to be $10^{-2}~$s and $10~$s, respectively, we estimate the infidelity caused by decoherence to be $<10^{-1}$ for $N_{\text{memories}}=10^3$ (see App.~\ref{sec:decoherence}). Engineering a $^{12}$C-rich environment~\cite{Jahnke_2012} could further improve the coherence times and thereby reduce the infidelity.

Other sources of infidelity include depolarization, measurement errors, and imperfect two-qubit interaction between nuclear and electron spins. To simplify the discussion, we combine all types of errors into one collective ``physical error rate'' $\epsilon$. We propose having interconnects interspersed between the layers that allow for arbitrary routing (see App.~\ref{sec:PIC_interconnect}). As a result, only the necessary number of nodes are activated and the teleportation scheme still adheres to the bucket-brigade model. Hence, the infidelity caused by physical errors is still $\epsilon^{\mathcal{O}(\log\left(N_{\text{memories}}\right))}$~\cite{Giovannetti_2008_PRL}.

However, for applications that require querying most addresses, the physical error rate could quickly decohere the qRAM since the infidelity rapidly grows as ${1-\mathcal{F}_q\propto (1-\epsilon)^{2^n-1}}$ for a circuit depth of $n$. Assuming a physical error rate of $\epsilon=10^{-4}$ and $n=10$, the query infidelity is already ${\sim 10^{-1}}$. Therefore, scaling up the qRAM necessitates further exploration in converting each tree node to a logical qubit and adapting quantum error correction~\cite{Nickerson_2014,Choi_2019}.

\section{Conclusion}

In summary, we introduced a qRAM implementation in a PIC platform integrated with solid-state spin memories. Our numerical simulations show that our architecture can achieve $>0.99$ fidelity with $>$kHz query rate for a qRAM containing $10^2$ memory cells. Moreover, our cavity-assisted scheme relies on heralding the requisite operations, thereby providing built-in qubit loss detection that further improves the query efficiency. Although high success rates demand a sufficiently over-coupled cavity to the waveguide, existing photonic crystal cavity designs~\cite{Quan_2010,Alajlan_2019,Vasco_2019} already show that they can reach near-unity coupling. We stress that our architecture is technologically feasible given rapidly advancing electro-optic platforms~\cite{Zhang_2017,Desiatov_2019} and experimentally shown large-scale integration of artificial atoms in PICs~\cite{Wan_2019}.

Additionally, we proposed an alternative scheme based on quantum teleportation that allows for efficiency scaling favorably with the circuit size. With sufficiently strong cavity-waveguide coupling, the teleportation approach enables $>$kHz query rate for a qRAM containing $10^5$ memories, a size unattainable by the conventional approach. We emphasize that the protocol is modular and can be applied to a quantum network, in which each network node acts as a tree node in the qRAM. The nodes would again be entangled via heralding, which removes qubit loss as a potential error.

The architecture also extends to other atomic memories: quantum dots~\cite{Sun_2016} and rare-earth ions~\cite{Kindem_2020} strongly coupled to nanocavities, and even trapped-ions~\cite{Pogorelov_2021} and neutral atoms~\cite{Omran_2019} suitable for creating large GHZ states. With rapid advancements in constructing high-fidelity atom-photon interfaces, our proposal presents a scalable design of a general qRAM in the NISQ era.

\section{Acknowledgements}

We thank Dr. Mikkel Heuck and Hyeongrak Choi for insightful discussions. K.C.C. acknowledges funding support by the National Science Foundation Graduate Research Fellowships Program (GRFP), the Army Research Laboratory Center for Distributed Quantum Information (CDQI), and the MITRE Corporation Moonshot program. W. Dai is supported by the National Science Foundation to the Computing Research Association for the CIFellows 2020 Program. C.E. acknowledges funding from the Swedish Research Council (2019-00684). This work was supported by the National Science Foundation QII-TAQS for Quantum Machine Learning with Photonics (1936314).

\newpage
\appendix
\setcounter{equation}{0}
\setcounter{figure}{0}
\renewcommand{\theequation}{S.\arabic{equation}}
\renewcommand{\thefigure}{S\arabic{figure}}
\renewcommand{\thetable}{S\arabic{table}}

\section{Atom-cavity parameters} \label{sec:atom_cavity}
The reflectivity of a single-sided cavity coupled with a quantum emitter is:
\begin{align}\label{eqn:supp:cav_reflectivity}
r(\omega) &= 1-\frac{\kappa_{\text{wg}}\left[i\Delta_a+\frac{\gamma}{2}\right]}{\left[i\Delta_c+\frac{\kappa}{2}\right]\left[i\Delta_c+\frac{\gamma}{2}\right]+g^2}
\end{align}
where $g$ is the atom-cavity coupling strength, $\gamma$ is the emitter's spontaneous emission rate, $\kappa$ is the cavity's total decay rate, $\kappa_{\text{wg}}$ is the waveguide-cavity coupling rate, and $\Delta_a=\omega_a-\omega$ and $\Delta_c=\omega_c-\omega$ are the atomic and cavity detuning from the probe, respectively. In the large cooperativity $C=4g^2/\kappa\gamma\gg 1$ limit and considering a perfectly over-coupled cavity, the reflectivity of an on-resonance probe $\Delta_a=\Delta_c=0$ simplifies to
\begin{align}
r(\omega) \xrightarrow{C\gg 1} \frac{C-1}{C+1}
\end{align}
Therefore, $r$ approaches $+1$ when $C$ increases, whereas a far off-resonance emitter decoupled from the cavity mode would yield $r\rightarrow-1$. In our cavity-assisted scheme, the photonic qubits are encoded in the frequency basis $\{\omega_0,\omega_1\}$. By appropriately choosing the atomic and cavity detuning, the resultant Fano interference can satisfy the following truth table, whose entry represents the probe's acquired phase from reflecting off the nanocavity:
\begin{table}[h!]
\centering
\begin{tabular}{c|c|c}
& $\ket{\downarrow}$ & $\ket{\uparrow}$\\
\hline
$\ket{\omega_0}$ & 0 & $\pi$\\
$\ket{\omega_1}$ & $\pi$ & 0
\end{tabular}
\end{table}

This can be satisfied by demanding the reflectivity to be $+1$ when the spin state is on-resonance and $-1$ when it is off-resonance. Using Eq.~\ref{eqn:supp:cav_reflectivity}, we arrive at the following equation:
\begin{align} \label{eqn:supp:cav_condition}
\text{Re}\left\{\frac{\kappa_{\text{wg}}\left(i\Delta_a+\frac{\gamma}{2}\right)}{\left(i\Delta_c+\frac{\kappa}{2}\right)\left(i\Delta_a+\frac{\gamma}{2}\right)+g^2}\right\} = 2
\end{align}

We let the cavity resonance centered at the average between the two transition frequencies: $\omega_c=(\omega_0+\omega_1)/2$. Therefore, given the Zeeman splitting $\Delta$, the cavity detuning would be half of the spin driving frequency: $\Delta_c = \frac{\Delta}{2}$. Similarly, the atomic detuning would exactly equal the splitting: $\Delta_a = \Delta$. In the Purcell regime, Eq.~\ref{eqn:supp:cav_condition} leads to the following condition:
\begin{align}
\Delta &\approx \sqrt{2g^2+\frac{\kappa}{4}\left(\kappa-\kappa_{\text{wg}}\right)-\frac{\gamma^2}{4}}
\end{align}

Therefore, given a fixed set of atom-cavity parameters $\{g,\gamma,\kappa,\kappa_{\text{wg}}\}$, we may set the corresponding magnetic field $B_{\text{opt}}$ that satisfies the appropriate Zeeman splitting $\Delta\sim\mu\text{g}B_{\text{opt}}/\hbar$ where $\mu=q\hbar/2m_e$ is the Bohr magneton and g$\approx 2$ is the Lande g-factor.

\begin{figure}[h!]
    \includegraphics[width=0.5\textwidth]{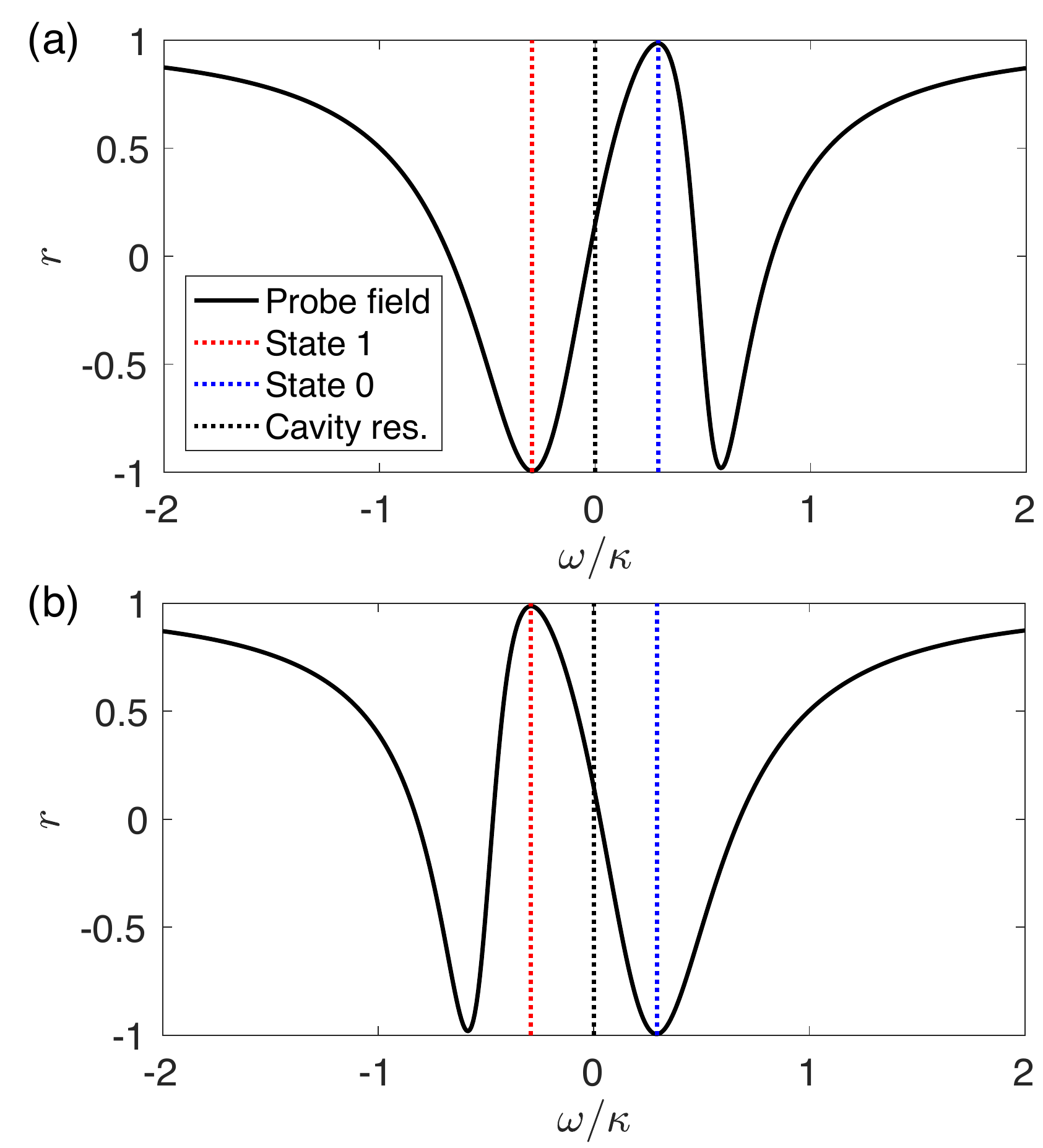}
    \caption{Cavity reflectivity as a function of probe frequency. The normalized probe frequency $\omega/\kappa$ is centered at the cavity resonance (black dashed line) $\omega_c$. The magnetic field is appropriately chosen such that the two atomic transition frequencies $\omega_0$ and $\omega_1$ coincide with the cavity reflectivity maximum $r=+1$ and minimum $r=-1$. The reflectivity when (a) the spin is in the $\ket{\downarrow}$ state is the mirror of when (b) the spin is in the $\ket{\uparrow}$ state.}
    \label{fig:supp:cav_reflectivity}
\end{figure}

As an illustrative example, we plot the reflectivity $r$ of a perfectly over-coupled cavity ($\kappa_{\text{wg}}/\kappa=1$) against the probe frequency $\omega/\kappa$. Fig.~\ref{fig:supp:cav_reflectivity}(a) shows $r=+1$ at the probe frequency $\omega=\omega_c+\Delta/2$ whereas $r=-1$ at $\omega=\omega_c-\Delta/2$ when the spin population resides in state $\ket{\downarrow}$, and vice versa as shown in Fig.~\ref{fig:supp:cav_reflectivity}(b). 

\section{Frequency-dependent add-drop filter} \label{sec:add_drop_filter}

\begin{figure}[h!]
    \centering
    \includegraphics[width=0.5\textwidth]{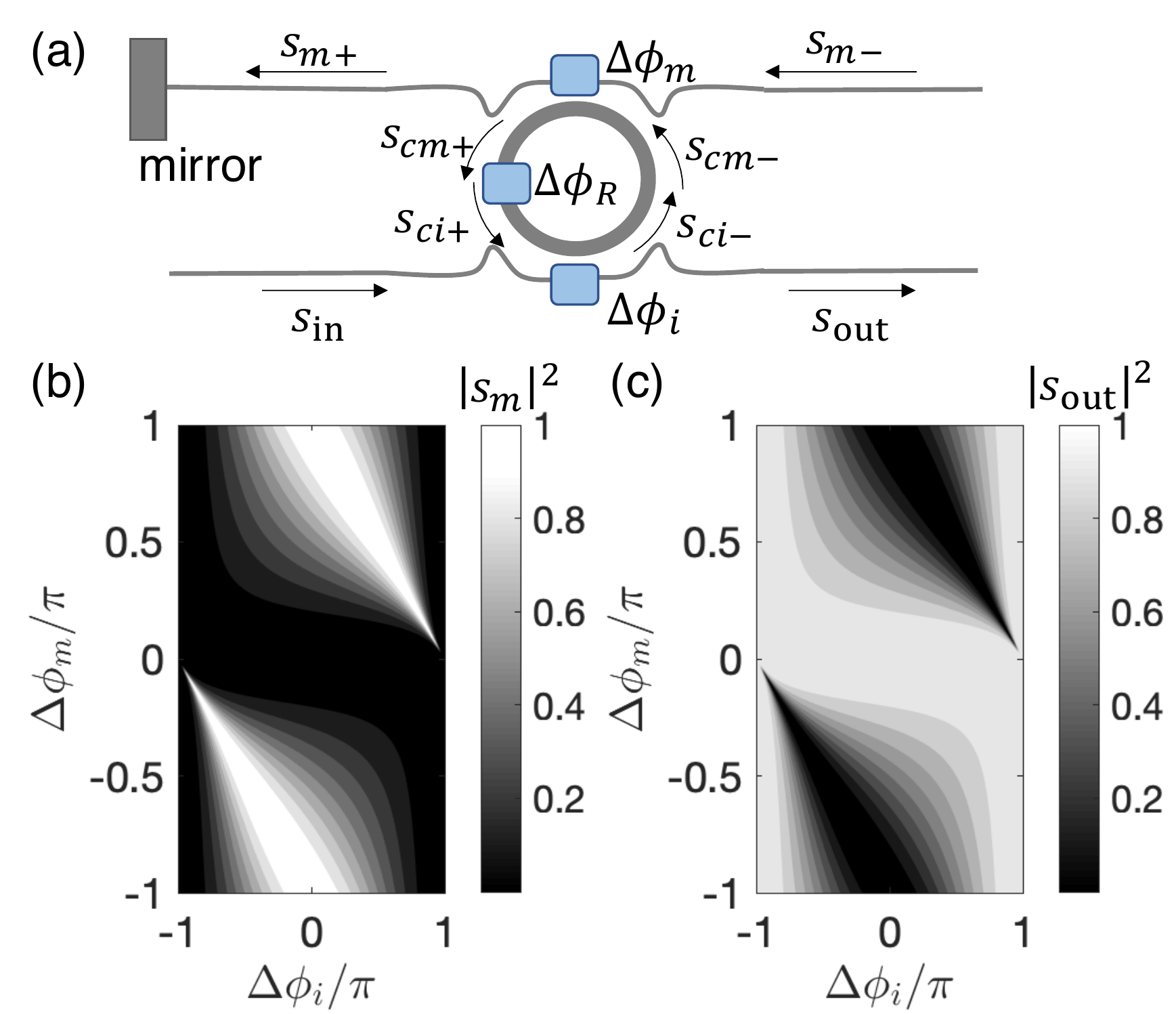}
    \caption{Add-drop filter schematic. (a) Each of the propagating fields in the add-drop filter is labeled for deriving the transfer matrices. The ring resonator (whose resonance can be tuned by $\Delta\phi_R$) is coupled to the waveguides via balanced MZI, or interferometric, couplers, each containing a phase shifter $\Delta\phi_{i,m}$. When the top waveguide is coupled to the resonator, the $\omega_0$ component is routed to reflect off a Sagnac loop reflector (mirror). (b) The output intensity towards the mirror $|s_m|^2$ as a function of $\Delta\phi_i$ and $\Delta\phi_m$. (c) The output intensity of the through-component $|\sout|^2$.}
    \label{fig:supp:sm_sout}
\end{figure}

To perform both the (1) setting and (2) routing operations, the add-drop filter must resonantly couple to only the $\omega_0$ component to impart (1) a $\pi$ phase shift upon reflection off an mirror and (2) a $\pi/2$ phase shift through a single pass after decoupling the resonator from the mirror waveguide. The system can be modeled by tracking the evolution of the field propagating through the MZI (or interferometric) couplers~\cite{Barbarossa_1995}. As illustrated in Fig.~\ref{fig:supp:sm_sout}(a), the outputs of the MZI couplers are:
\begin{align}
\left[\begin{matrix}
\sout\\s_{ci-}
\end{matrix}\right] &= \mathcal{T}^{(i)}\left[\begin{matrix}
\ssin\\s_{ci+}
\end{matrix}\right], \quad \left[\begin{matrix}
s_{m+}\\s_{cm+}
\end{matrix}\right] = \mathcal{T}^{(m)}\left[\begin{matrix}
s_{m-}\\s_{cm-}
\end{matrix}\right]
\end{align}
where $\mathcal{T}^{(n)}=C^{(n)}Z^{(n)}C^{(n)}$ for $n=\{m,i\}$. The matrices $C^{(n)}$ and $Z^{(n)}$ are transfer matrices that describe the beam splitter and the interferometer arms:
\begin{align}
C^{(n)} &= \left[\begin{matrix}
\nu_n & i\sqrt{1-\nu_n^2}\\
i\sqrt{1-\nu_n^2} & \nu_n
\end{matrix}\right], Z^{(n)} &= \left[\begin{matrix}
e^{i\Psi_{nT}} & 0\\
0 & e^{i\Psi_{nB}}
\end{matrix}\right]
\end{align}
where $\nu_n$ represents the coupling to the through-waveguide, $\Psi_{nT}$ and $\Psi_{nB}$ are the phases accumulated in the phase shifter and the resonator arms, respectively. For the remainder of the section, we assume a balanced interferometric coupler such that $\nu_n=1/\sqrt{2}$.

Explicitly, we can write the MZI transfer matrix as:
\begin{align}
\mathcal{T} &= e^{i\Psi_{nR}}\left[\begin{matrix}
(1+e^{i\phi_n})\nu_n^2-1 & i(1+e^{i\phi_n})\nu_n\sqrt{1-\nu_n^2}\\
i(1+e^{i\phi_n})\nu_n\sqrt{1-\nu_n^2} & \nu_n^2-e^{i\phi_n}(1-\nu_n^2)
\end{matrix}\right]\nonumber\\
&\quad\forall n\in\{ m,i\}
\end{align}
where $\phi_n(\omega)=k(\omega)\Delta L_n+\Delta\phi_n$ and $k(\omega)=(\neff/c)\omega_0+(n_{g,\text{PIC}}/c)(\omega-\omega_0)$. Here, $\Delta L_n$ is the path length difference between the two arms and $k(\omega)$ is the propagation constant governed by the effective and group indices in the PIC, $\neff$ and $n_{g,\text{PIC}}$, respectively.

For the interest of our operations, we can set $\ssin=1$ and $S_{m-}=0$. The resultant system of equations is:
\begin{align}
\sout &= T_{1,1}^{(i)} \ssin+T_{1,2}^{(i)}s_{ci+}\\
s_{ci-} &= T_{2,1}^{(i)} \ssin+ T_{2,2}^{(i)}s_{ci+}
\end{align}
From which, after solving for $\sout$ and $s_{m+}=T_{1,2}^{(m)}s_{cm-}$, we get:
\begin{align}
\sout &= e^{i\Psi_{iR}}\left(T_{1,1}^{'(i)}+\frac{e^{i\phi_c}\zeta_m T_{1,2}^{'(i)}T_{2,1}^{'(i)}}{1-e^{i\phi_c}\zeta_i\zeta_m}\right)\\
s_{m+} &= \frac{e^{i\phi_{im}}T_{1,2}^{(m)}T_{2,1}^{(i)}}{1-e^{i\phi_c}\zeta_i\zeta_m}
\end{align}
where $\phi_c(\omega)=\psi_{iR}+\phi_{im}+\psi_{mR}+\phi_{mi}=k(\omega)L_c$ is the phase acquired in the resonator, and $\zeta_n=\nu_n^2-e^{i\phi_n}(1-\nu_n^2)$. For the routing operation, we wish to have $s_m=s_{m+}=1$ (correspondingly $\sout=0$) such that the $\omega_0$ component is entirely directed to the mirror. In Fig.~\ref{fig:supp:sm_sout}(b,c), we plot the output intensity $|s_m|^2$ and $|\sout|^2$ as a function of $\Delta\phi_i$ and $\Delta\phi_m$ set by the phase shifters in the MZI couplers. In order to maximize $|s_m|^2$, we find that the phases must satisfy the condition: $\Delta\phi_i+\Delta\phi_m=\pi$. 

\begin{figure}[h!]
    \centering
    \includegraphics[width=0.5\textwidth]{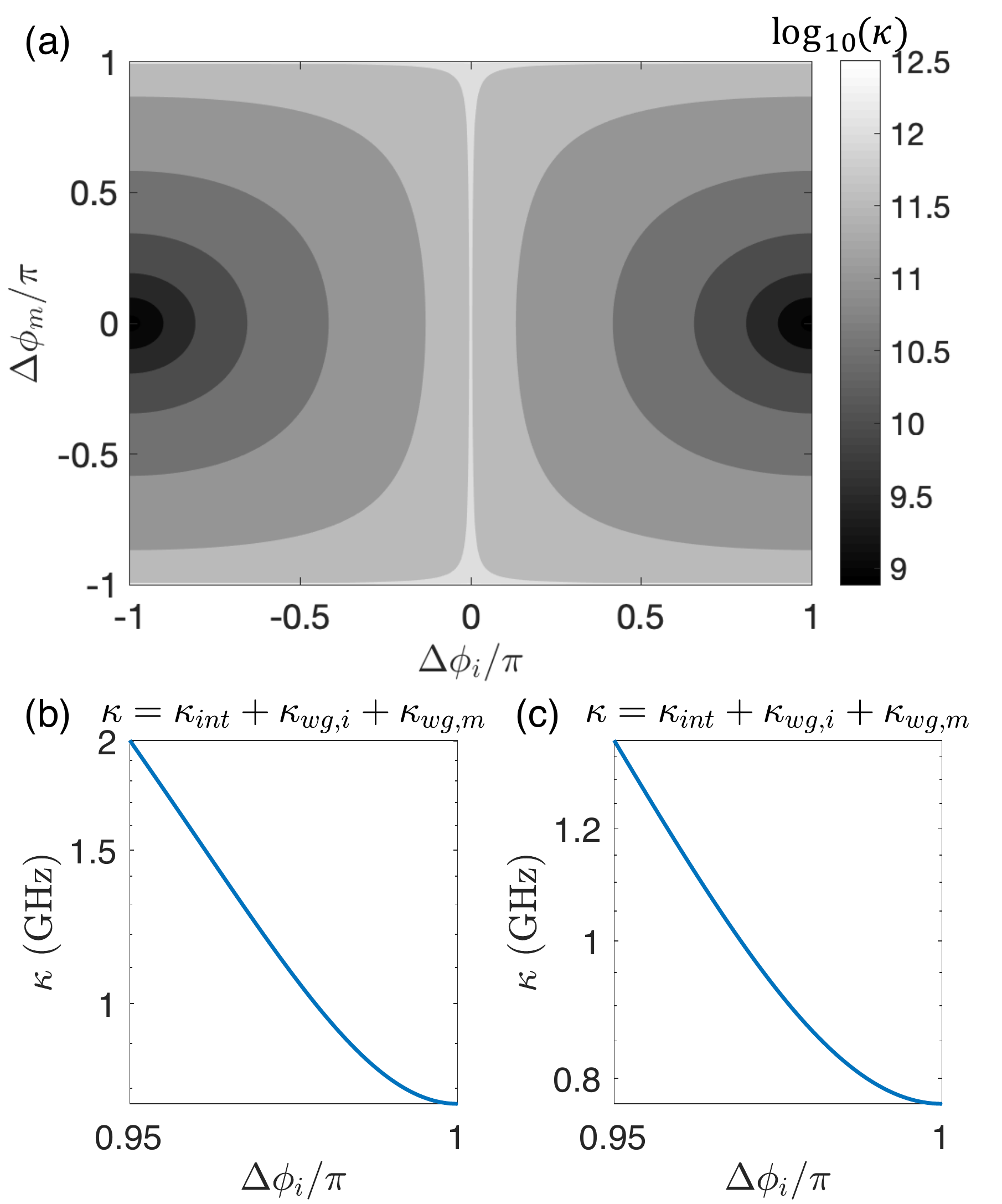}
    \caption{Decay rate of the ring resonator. (a) The resonator's total decay rate (linewidth) is plotted as a function of $\Delta\phi_i$ and $\Delta\phi_m$ on a log scale. $\kappa$ reaches its minimum near $\Delta\phi_i=\pm\pi$ and $\Delta\phi_m=0$ at which the resonator is decoupled from the waveguides. $\kappa$ (GHz) is plotted against $\Delta\phi_i$ for (b) the setting mode and (c) the routing mode.}
    \label{fig:supp:kappa}
\end{figure}

It is equally essential for the resonator to have a sufficiently high quality factor (Q) such that the linewidth is narrow enough to only couple to the $\omega_0$ instead of both frequencies. For the simulations presented in the main text, the Zeeman splitting is assumed to be $\sim 12~$GHz, which implies that the Q must be $>10^4$ to resolve between $\omega_0$ and $\omega_1$. In Fig.~\ref{fig:supp:kappa}(a), we find that $\kappa$ is smallest at $\Delta\phi_i=\pm\pi$, which corresponds to the resonator completely decoupled from the input waveguide (source) and cavity leakage is maximally suppressed. Similarly, when $\Delta\phi_m=0$, the ring (source) is completely decoupled from the mirror waveguide. As long as $\Delta\phi_i$ is sufficiently close to $\pi$, Fig.~\ref{fig:supp:kappa}(b) indicates that the resonator linewidth is sufficiently smaller than the Zeeman splitting of $\sim 12~$GHz. For example, at $\Delta\phi_i=0.95\pi$ such that $\Delta\phi_m=0.05\pi$, $|s_m|^2$ is approximately unity and hence satisfies the setting mode. In the routing mode, we only need to minimally shift $\Delta\phi_m$ to 0 such that $|\sout|^2=1$ and $|s_m|^2=0$, as indicated by the drastically varying region near $\Delta\phi_i=\pi$ and $\Delta\phi_m=0$, as shown in Fig.~\ref{fig:supp:sm_sout}(b,c). With $\Delta\phi_m=0$ fixed, we validate that the narrowness of the resonator linewidth as illustrated by Fig.~\ref{fig:supp:kappa}(c). $\kappa$ is expectedly smaller in the routing mode than the setting mode since the resonator is decoupled from the mirror waveguide, thereby having one fewer leakage channel.

Lastly, we can appropriately choose $\Delta\phi_R$, which is the phase shifter within the resonator, such that traversing through the resonator imparts a $\pi/2$ phase to the $\omega_0$ component upon a single pass. In a round-trip, $\ket{\omega_0}$ effectively undergoes a Pauli $X$ gate, rendering the truth table:
\begin{table}[h!]
\centering
\begin{tabular}{c|c|c}
& $\ket{\downarrow}$ & $\ket{\uparrow}$\\
\hline
$\ket{\omega_0}$ & 0 & $\pi$\\
$\ket{\omega_1}$ & $\pi$ & 0
\end{tabular} $\xrightarrow[X\ \text{on}\ \omega_0]{}$
\begin{tabular}{c|c|c}
& $\ket{\downarrow}$ & $\ket{\uparrow}$\\
\hline
$\ket{\omega_0}$ & $\pi$ & 0\\
$\ket{\omega_1}$ & $\pi$ & 0 
\end{tabular} 
\end{table}

\section{Quantum state transfer} \label{sec:state_transfer}
\subsection{Photon-to-spin}
The atom is first initialized in a superposition of the two ground states: $\ket{\psi_A}=(\ket{\downarrow}+\ket{\uparrow})/\sqrt{2}$. With the incoming frequency-encoded photonic qubit, $\ket{\psi_P} = \alpha\ket{\omega_0}+\beta\ket{\omega_1}$, the joint (unnormalized) photon-atom state is:
\begin{align}
\ket{\psi} &= \ket{\psi_P}\otimes \ket{\psi_A}\nonumber\\
&= (\alpha\ket{\omega_0}+\beta\ket{\omega_1})(\ket{\downarrow}+\ket{\uparrow})
\end{align}

The add-drop filter resonantly couples to only the $\omega_0$ component that then reflects off a mirror, acquiring $\pi$ phase shift regardless of the atomic state. On the other hand, the $\omega_1$ component interacts with the atom-cavity system and acquires a spin-dependence phase shift. After the CZ operation, the photon and the atom are entangled:
\begin{align}
\ket{\psi} &=-\alpha\ket{\omega_0,\downarrow}-\alpha\ket{\omega_0,\uparrow}-\beta\ket{\omega_1,\downarrow}+\beta\ket{\omega_1,\uparrow}
\end{align}

The returning photon then goes through a frequency beam splitter that performs a Hadamard gate. After which, the two frequency components are routed to different photon detectors:
\begin{align}
\ket{\psi} &=-\alpha(\ket{\omega_0}+\ket{\omega_1})(\ket{\downarrow}+\ket{\uparrow})\nonumber\\
&\quad -\beta(\ket{\omega_0}-\ket{\omega_1})(\ket{\downarrow}-\ket{\uparrow})\nonumber\\
&= \ket{\omega_0} \otimes \left[ -(\alpha+\beta)\ket{\downarrow}-(\alpha-\beta)\ket{\uparrow}\right]\nonumber\\
&\quad +\ket{\omega_1}\otimes\left[-(\alpha-\beta)\ket{\downarrow}-(\alpha+\beta)\ket{\uparrow}\right]
\end{align}

Upon heralding, the atom undergoes another Hadamard gate to complete quantum teleportation:
\begin{align}
\ket{\psi} &= \ket{\omega_0} \otimes \left[ -(\alpha+\beta)(\ket{\downarrow}+\ket{\uparrow})-(\alpha-\beta)(\ket{\downarrow}-\ket{\uparrow})\right]\nonumber\\
&\quad +\ket{\omega_1}\otimes\left[-(\alpha-\beta)(\ket{\downarrow}+\ket{\uparrow})-(\alpha+\beta)(\ket{\downarrow}-\ket{\uparrow})\right]\nonumber\\
&= -\ket{\omega_0}\otimes(\alpha\ket{\downarrow}+\beta\ket{\uparrow})+\ket{\omega_1}\otimes(-\alpha\ket{\downarrow}+\beta\ket{\uparrow})
\end{align}

The end result is:
\begin{align*}
\ket{\psi} &= \alpha\ket{\downarrow}+\beta\ket{\uparrow}\quad \text{if}\ \omega_0\ \text{is detected}\nonumber\\
&\text{or}\quad \alpha\ket{\downarrow}-\beta\ket{\uparrow}\quad \text{if}\ \omega_1\ \text{is detected}
\end{align*}
neglecting global phase. Note that an additional Pauli-$Z$ operation is needed if $\omega_1$ is detected.

Now, let us consider an imperfectly over-coupled single-sided cavity with waveguide-cavity coupling $\kappa_{\text{wg}}/\kappa<1$. We denote $\roff$ and $\ron$ as the off- and on-resonance cavity reflectivities, and $r_m$ as the mirror reflectivity. Assuming the interferometric couplers are lossless in the add-drop filter, the photon-atom entangled state is then:
\begin{align}
\ket{\psi} &=\alpha r_m\ket{\omega_0,\downarrow}+\alpha r_m\ket{\omega_0,\uparrow}+\beta \roff\ket{\omega_1,\downarrow}+\beta \ron\ket{\omega_1,\uparrow}
\end{align}

After the Hadamard on the photon:
\begin{align}
\ket{\psi} &= \alpha r_m(\ket{\omega_0}+\ket{\omega_1})(\ket{\downarrow}+\ket{\uparrow})\nonumber\\
&\quad +\beta(\ket{\omega_0}-\ket{\omega_1})(\roff\ket{\downarrow}+\ron\ket{\uparrow})\nonumber\\
&= \ket{\omega_0} \otimes \left[ (\alpha r_m +\beta \roff)\ket{\downarrow}+(\alpha r_m+\beta \ron)\ket{\uparrow}\right]\nonumber\\
&\quad +\ket{\omega_1}\otimes\left[(\alpha r_m-\beta \roff)\ket{\downarrow}+(\alpha r_m-\beta \ron)\ket{\uparrow}\right]
\end{align}

The additional Hadamard on the atom would yield:
\begin{align}
\ket{\psi} &= \ket{\omega_0} \otimes \left[ (\alpha r_m+\beta \roff)(\ket{\downarrow}+\ket{\uparrow})\right.\nonumber\\
&\quad\quad \left.+(\alpha r_m+\beta \ron)(\ket{\downarrow}-\ket{\uparrow})\right]\nonumber\\
&\quad+\ket{\omega_1}\otimes\left[(\alpha r_m-\beta \roff)(\ket{\downarrow}+\ket{\uparrow})\right.\nonumber\\
&\quad\quad\left.+(\alpha r_m-\beta \ron)(\ket{\downarrow}-\ket{\uparrow})\right]\nonumber\\
&= \ket{\omega_0} \otimes \left[ (2\alpha r_m+\beta(\ron+\roff))\ket{\downarrow}\right.\nonumber\\
&\quad\quad\left. +\beta(-\ron+\roff)\ket{\uparrow}\right]\nonumber\\
&\quad +\ket{\omega_1}\otimes\left[(2\alpha r_m-\beta(\ron+\roff))\ket{\downarrow}\nonumber\right.\\
&\quad\quad\left.+\beta(\ron-\roff)\ket{\uparrow}\right]
\end{align}

If the register qubit is $\ket{\psi_P}=(\ket{\downarrow}+\ket{\uparrow})/\sqrt{2}$ such that $\alpha=\beta=1/\sqrt{2}$ and we assume $|r_m|=1$, detection on the $\omega_0$ port would herald the state:
\begin{align}
\ket{\psi} &= (2+\ron+\roff)\ket{\downarrow}+(-\ron+\roff)\ket{\uparrow}
\end{align}

Since $\text{sgn}(\ron)=1$ and $\text{sgn}(\roff)=-1$, we see that $\ket{\psi}\Rightarrow\ket{\downarrow}+\ket{\uparrow}$ requires $|\ron|=|\roff|$, which hints at the need to ``balance'' these two reflectivities. Eq.~\ref{eqn:supp:cav_reflectivity} dictates that $\ron\propto \kappa_{\text{wg}}(C-1)/(C+1)$ while $\roff\propto\kappa_{\text{wg}}/\kappa$ such that only a suitable regime of $\{g,\gamma,\kappa,\kappa_{\text{wg}}\}$ would maximize the quantum state transfer fidelity as shown in Fig.~\ref{fig:fidelity}.

\subsection{Spin-to-photon}
Once the bus qubit retrieves the data from the memory layer, we must extract the address out of the qRAM to obtain the correlated output state $\sum_j\alpha_j\ket{j}_a\ket{D_j}_b$. By sending additional photons, we can perform quantum state transfer that \textit{maps the spin qubits onto the photonic qubits}. Similar to the heralding procedure for transferring the photonic states to spin qubits, the spins must undergo projective measurements to complete the spin-to-photon mapping. While it is feasible to perform single shot readout on one spin, it is experimentally difficult to simultaneously perform projective measurements on \textit{multiple spins} within one layer. The issue can be circumvented by introducing an ancillary photon that is entangled with the spins for each layer, and heralding on such photon equates to performing projective readout on the spin qubits.

After data retrieval, the spin holds the routing state $\ket{\psi_A}=\alpha\ket{\downarrow}+\beta\ket{\uparrow}$. The incoming photon initialized in the superposition state (un-normalized) $\ket{\psi_{P1}}=\ket{\omega_0}_1+\ket{\omega_1}_1$ interacts with the cavity, producing the output state:
\begin{align}
\ket{\Psi} &= -\alpha\ket{\omega_0}_1\ket{\downarrow}-\alpha\ket{\omega_1}_1\ket{\downarrow}-\beta\ket{\omega_0}_1\ket{\downarrow}+\beta\ket{\omega_1}_1\ket{\uparrow}\nonumber\\
&= -\alpha(\ket{\omega_0}_1+\ket{\omega_1}_1)\ket{\downarrow}-\beta(\ket{\omega_0}_1-\ket{\omega_1}_1)\ket{\uparrow}
\end{align}

After a Hadamard operation on the spin qubit, the entangled state becomes:
\begin{align}
\ket{\Psi} &= -\alpha(\ket{\omega_0}_1+\ket{\omega_1}_1)(\ket{\downarrow}+\ket{\uparrow})\nonumber\\
&\quad -\beta(\ket{\omega_0}_1-\ket{\omega_1}_1)(\ket{\downarrow}-\ket{\uparrow})
\end{align}

A subsequent Hadamard operation (via the frequency beam splitter) on the photon yields:
\begin{align}
\ket{\Psi} &= -\alpha\ket{\omega_0}_1(\ket{\downarrow}+\ket{\uparrow}) -\beta\ket{\omega_1}_1(\ket{\downarrow}-\ket{\uparrow})\nonumber\\
&= -\ket{\downarrow}\otimes (\alpha\ket{\omega_0}_1+\beta\ket{\omega_1}_1)-\ket{\uparrow}\otimes(\alpha\ket{\omega_0}_1-\beta\ket{\omega_1}_1)
\end{align}

We then send a subsequent photon $\ket{\Psi_{P2}}=\ket{\omega_0}_2+\ket{\omega_1}_2$ that will entangle with the spin qubit for performing the projective measurement. Similarly, the composite state undergoes a CZ operation upon cavity reflection, resulting in:
\begin{align}
\ket{\Psi} &= \ket{\downarrow}(\ket{\omega_0}_2+\ket{\omega_1}_2) (\alpha\ket{\omega_0}_1+\beta\ket{\omega_1}_1)\nonumber\\
&\quad +\ket{\uparrow}(\ket{\omega_0}_2-\ket{\omega_1}_2)(\alpha\ket{\omega_0}_1-\beta\ket{\omega_1}_1)
\end{align}

Another Hadamard operation on the second photon would produce an entangled state:
\begin{align}
\ket{\Psi} &= \ket{\downarrow}\ket{\omega_0}_2(\alpha\ket{\omega_0}_1+\beta\ket{\omega_1}_1)\nonumber\\
&\quad +\ket{\uparrow}\ket{\omega_1}_2(\alpha\ket{\omega_0}_1-\beta\ket{\omega_1}_1)
\end{align}

As a result, any projection on the frequency-encoded photon is a projective measurement on the spin as well. If $\ket{\omega_0}_2$ is detected, the effective projection onto $\ket{\downarrow}$ results in the transferred state onto the first photon. Instead, if $\ket{\omega_1}_2$ is detected, an additional $\pi$-pulse would be applied to the first photon to construct $\alpha\ket{\omega_0}_1+\beta\ket{\omega_1}_1$. Imperfections in the cavity system would be treated in the same fashion as the previous section by taking account non-unity reflectivities: $\ron,\roff,r_m$.

\section{Quantum routing} \label{sec:quantum_routing}
In the routing mode, the MZI in Fig.~\ref{fig:operations}(e) is tuned to operate as a 50:50 beam splitter whose unitary matrix is denoted as $B$. Let $a,b$ be the annihilation operators for the top and bottom spatial modes such that $\adag\ket{0}_a\ket{0}_b=\ket{1}_a\ket{0}_b$ represents one photon present in the top waveguide and no photon in the bottom waveguide. The MZI provides the following unitary transformation on the operators:
\begin{align}
Ba\bsdag &= \frac{1}{\sqrt{2}} (a+ib)\\
Bb\bsdag &= \frac{1}{\sqrt{2}} (b+ia)
\end{align}

Assuming input from strictly the top waveguide, our initial state is $\ket{\phi_0}=\ket{1}_a\ket{0}_b=\adag\ket{0}_a\ket{0}_b$. After passing through the MZI, the state becomes:
\begin{align}
\ket{\phi_1} &= B\ket{\phi_0}\nonumber\\
&= B\adag\ket{0}_a\ket{0}_b = B\adag\bsdag B\ket{0}_a\ket{0}_b\nonumber\\
&= \frac{1}{\sqrt{2}}(\adag-i\bdag)\ket{0}_a\ket{0}_b
\end{align}

At this point, recall that the photon exiting out of the bottom output inherits a $\pi$ phase shift upon a reflection off the mirror. The photon exiting out of the top output receives either no phase shift or a $\pi$ phase shift depending on the spin state. Therefore, the cavity system effectively acts a phase shifter controlled by the spin qubit. We can denote the unitary transformation of the atom-coupled cavity system (in conjunction with the resonator) as $PaP^{\dagger}=ae^{i\phi}$. Note that it is only acting on the top waveguide and has no effect on $b$. As a result, the photonic qubit after reflection off the mirror and the cavity system becomes:
\begin{align}
\ket{\phi_2} &= P\ket{\phi_1}\nonumber\\
&= \frac{1}{\sqrt{2}}(P\adag-i\bdag)P^{\dagger}P\ket{0}_a\ket{0}_b\nonumber\\
&= \frac{1}{\sqrt{2}}(e^{i\phi}\adag-i\bdag)\ket{0}_a\ket{0}_b
\end{align}

Lastly, the photon returns to and interacts with the MZI once again:
\begin{align}
\ket{\phi_3} &= \bsdag\ket{\phi_2}\nonumber\\
&= \frac{1}{\sqrt{2}}\bsdag(e^{i\phi}\adag-i\bdag)B\bsdag\ket{0}_a\ket{0}_b\nonumber\\
&= \frac{1}{2}\left( e^{i\phi}(\adag+i\bdag)-i(\bdag+i\adag)\right)\ket{0}_a\ket{0}_b\nonumber\\
&= e^{i\phi/2}\left[\left(\frac{e^{i\phi/2}+e^{-i\phi/2}}{2}\right)\adag\right.\nonumber\\
&\quad\left.+i\left(\frac{e^{i\phi/2}-e^{-i\phi/2}}{2}\right)\bdag\right]\ket{0}_a\ket{0}_b\nonumber\\
&= e^{i\phi/2}\left[\cos\left(\frac{\phi}{2}\right)\ket{1}_a\ket{0}_b-\sin\left(\frac{\phi}{2}\right)\ket{0}_a\ket{1}_b\right]
\end{align}

If the spin is in the down state $\ket{\downarrow}$, the photon acquires a $\pi$ phase shift. However, because the mirror reflection imparts a $\pi$ phase shift as well, the relative phase $\phi$ between the two arms is effectively zero. Hence, the output would all go to the top path, i.e. the spatial mode $\ket{1}_a$. On the other hand, if the spin is in the up state $\ket{\uparrow}$, a $\pi$ relative phase is acquired in the interferometer, resulting in an output going to the bottom path, i.e. the spatial mode $\ket{1}_b$.

\section{PIC implementation} \label{sec:PIC_implementation}

\begin{figure}[h!]
    \centering
    \includegraphics[width=0.5\textwidth]{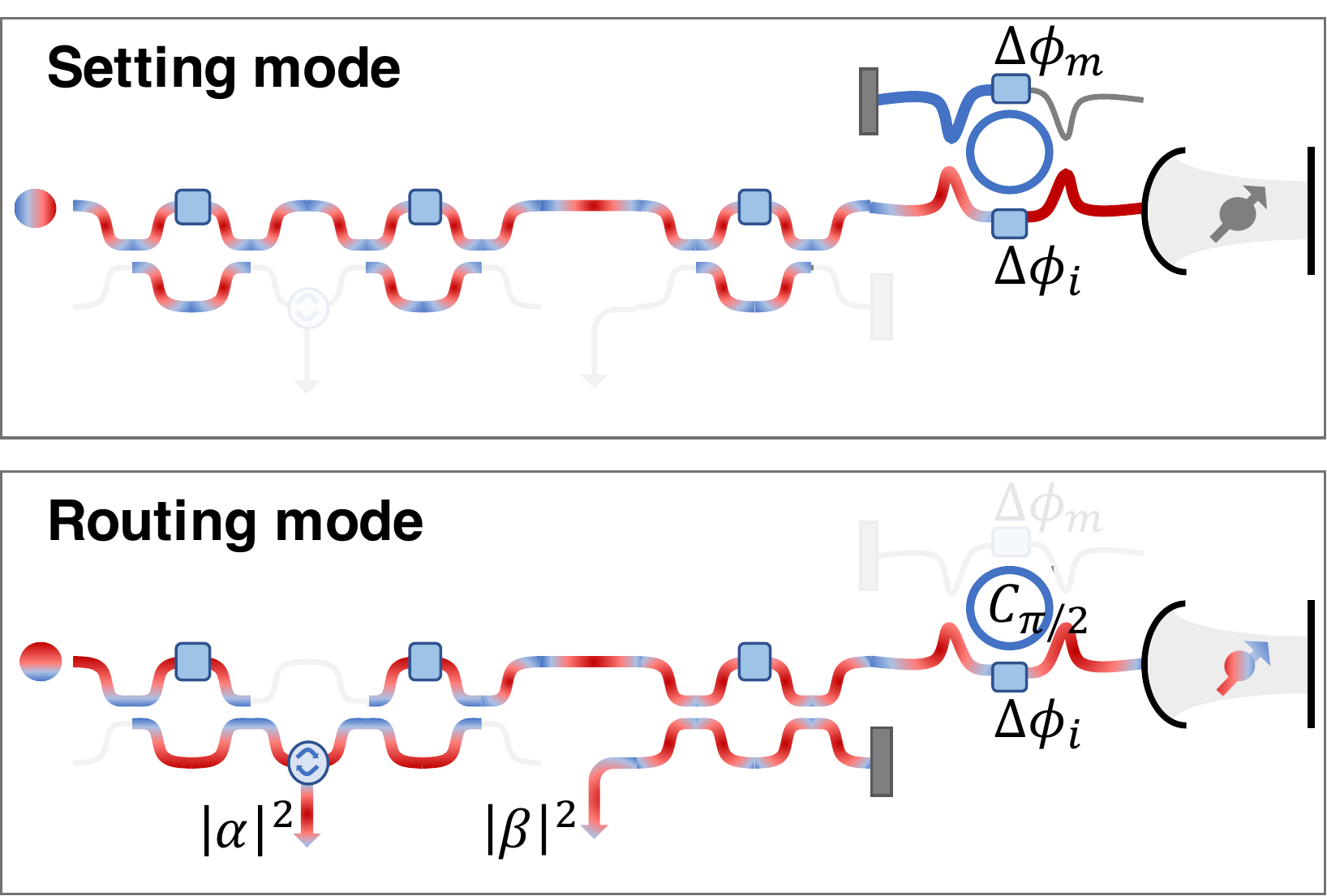}
    \caption{A detailed PIC implementation of a tree node. Paths that are inactive in each mode are faded out.}
    \label{fig:supp:implementation}
\end{figure}

Fig.~\ref{fig:supp:implementation} illustrates a more detailed schematic of our PIC implementation for each tree node. Specifically, a circulator is appended so the incoming photon can be routed to the children nodes as opposed to returning to the root.

\section{Success probability} \label{sec:success_probability}

\begin{figure}[h!]
    \centering
    \includegraphics[width=0.5\textwidth]{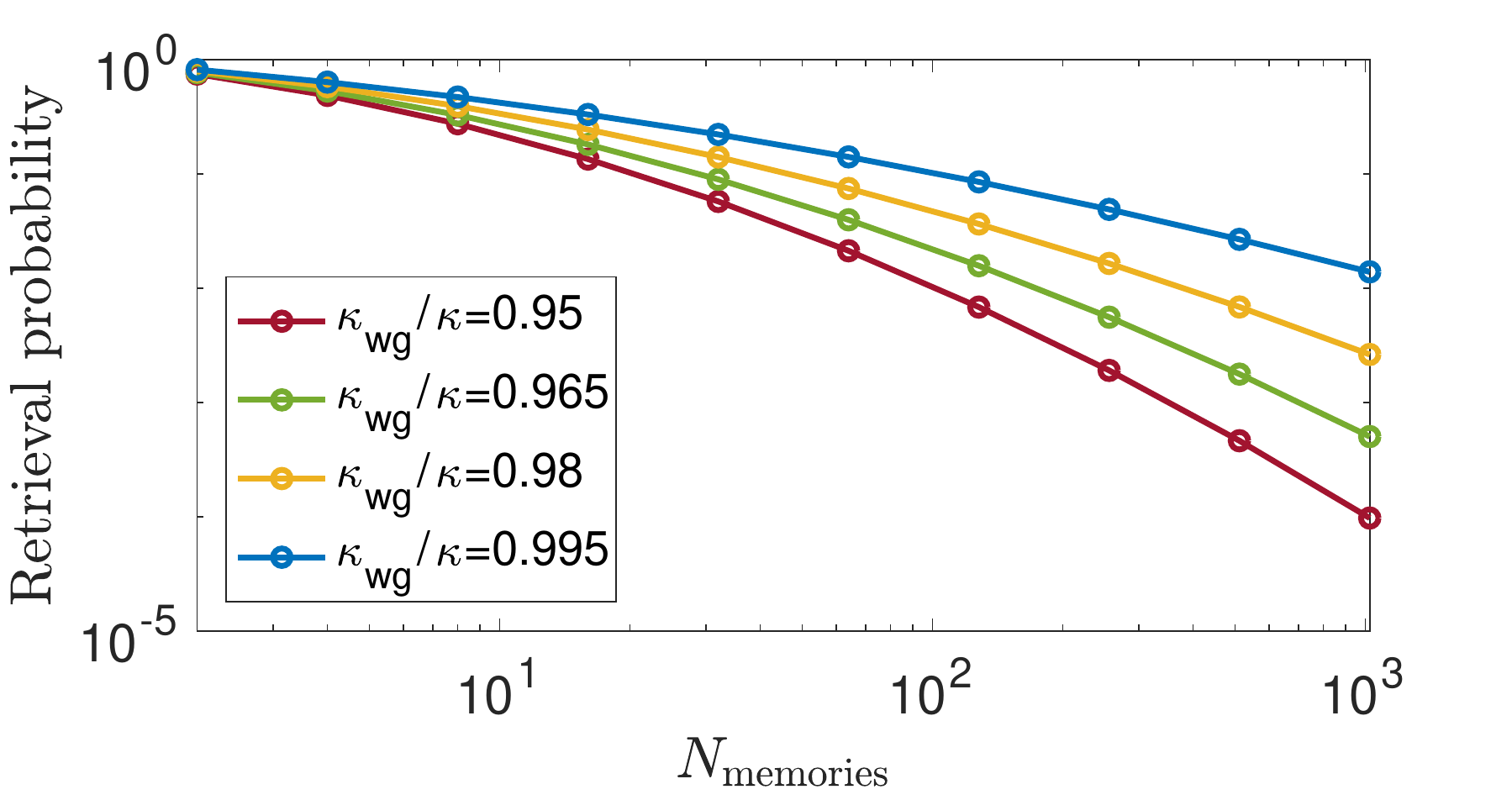}
    \caption{The retrieval probability as a function of the number of memories on a log-log scale. $p_{\text{succ}}$ displays a polynomial roll-off as $N_{\text{memories}}$ increases for waveguide-cavity coupling $\kappa_{\text{wg}}/\kappa$=0.95,0.965,0.98,0.995.}
    \label{fig:supp:layer_probability}
\end{figure}

For the bus qubit to reach the memory layer in a $n$-level qRAM, each register photon must traverse to layer $i<n$. Therefore, considering a propagation loss of $\eta_p$ (see Table~\ref{tab:parameters}), the probability of reaching layer $i$ is $e^{-\eta_p L(i)}$, where $L(i)$ is the distance between the said layer and the root node. Since the photon can scatter off the single-sided cavity and the mirror, interaction at each layer reduces the probability of detecting a returning register photon by $R_{\text{cav}}$ and $R_m$, which represent the cavity and mirror reflections, respectively. If we define the setting efficiency as $\eta_s=\eta_{\text{det}}(R_{\text{cav}}+R_m)/2$ and the routing efficiency as $\eta_r=R_{\text{cav}}$, the probability of completing each layer $i$ is then:
\begin{align}
p_i &= e^{-\eta_p L(i)}\eta_r^{i-1}\eta_s \quad \text{for}\ i\in\{1,...,n\}
\end{align}

A successful qRAM query would consequently occur with a probability that is the product of all the layer probabilities:
\begin{align}
p_{\text{succ}}&=\prod_{i=1}^n p_i = e^{-\sum_i\eta_p(i)L(i)}\eta_r^{n(n-1)/2}\eta_s^n
\end{align}

Expectedly, Fig.~\ref{fig:supp:layer_probability} shows a polynomial roll-off in the success probability $p_{\text{succ}}$ as the number of memory cells $N_{\text{memories}}=2^n$ increases. Since the waveguide-cavity coupling $\kappa_{\text{wg}}/\kappa$ mainly determines the cavity reflection, $p_{\text{succ}}$ can differ by orders of magnitude and the difference increases with $N_{\text{memories}}$.

\section{Teleportation scheme} \label{sec:teleportation}

Essential to the setup of the teleportation scheme is to create a GHZ state for each layer prior to quantum teleportation. Below, we break down its creation process into 3 critical steps: photon-assisted Bell state creation, Bell state swap between nuclear (memory) and electron (broker) spins, and GHZ state creation by joining adjacent pairs. After which, we explain how a Bell state measurement can be made on two remotely entangled spins via the photon-assisted cavity interaction. Lastly, we provide an example of how teleportation enables transferring addresses onto the qRAM.

\subsection{Photon-assisted Bell state creation} \label{sec:bell_state_creation}

In order to create a Bell state between neighboring matter qubits, a photon is sent to reflect off each cavity consecutively. Importantly, the node is in the ``setting'' mode such that reflection off the cavity system generates a CZ gate. Here, we provide an example of how a photon interacting with two cavities aids construction of a Bell state between the two spin qubits. We begin with the photonic and the spin qubits prepared in the $\ket{+}$ state such that composite state is:
\begin{align}
\ket{\psi} &= \left(\ket{\omega_0}+\ket{\omega_1}\right)\left(\ket{\downarrow}_1+\ket{\uparrow}_1\right)\left(\ket{\downarrow}_2+\ket{\uparrow}_2\right)
\end{align}
where the subscripts 1 and 2 denote different spins.

After the photon reflects off the first spin qubit coupled to the cavity, the state becomes an entangled state:
\begin{align}
\ket{\psi} &= -\left[\left(\ket{\omega_0}+\ket{\omega_1}\right)\ket{\downarrow}_1+\left(\ket{\omega_0}-\ket{\omega_1}\right)\ket{\uparrow}_1\right]\left(\ket{\downarrow}_2+\ket{\uparrow}_2\right)\nonumber\\
&= -\left[\left(\ket{\omega_0}\ket{\downarrow}_2+\ket{\omega_1}\ket{\downarrow}_2+\ket{\omega_0}\ket{\uparrow}_2+\ket{\omega_1}\ket{\uparrow}_2\right)\ket{\downarrow}_1\right.\nonumber\\
&\quad +\left.\left(\ket{\omega_0}\ket{\downarrow}_2-\ket{\omega_1}\ket{\downarrow}_2+\ket{\omega_0}\ket{\uparrow}_2-\ket{\omega_1}\ket{\uparrow}_2\right)\ket{\uparrow}_1\right]
\end{align}

Upon reflecting off the second cavity system, it produces the state:
\begin{align}
\ket{\psi} &= \left[\left(\left(\ket{\omega_0}+\ket{\omega_1}\right)\ket{\downarrow}_2+\left(\ket{\omega_0}-\ket{\omega_1}\right)\ket{\uparrow}_2\right)\ket{\downarrow}_1\right.\nonumber\\
&\quad +\left.\left(\left(\ket{\omega_0}-\ket{\omega_1}\right)\ket{\downarrow}_2+\left(\ket{\omega_0}+\ket{\omega_1}\right)\ket{\uparrow}_2\right)\ket{\uparrow}_1\right]
\end{align}

A Hadamard operation on the photon leads to the final state:
\begin{align}
\ket{\psi} &= \ket{\omega_0}\left(\ket{\downarrow\downarrow}+\ket{\uparrow\uparrow}\right)+\ket{\omega_1}\left(\ket{\downarrow\uparrow}+\ket{\uparrow\downarrow}\right)
\end{align}
If the $\omega_0$ detection port clicks, the Bell state $\ket{\Phi^+}=\ket{\downarrow\downarrow}+\ket{\uparrow\uparrow}$ is heralded. On the other hand, if the $\omega_1$ port registers a click, the Bell state $\ket{\Psi^+}=\ket{\downarrow\uparrow}+\ket{\uparrow\downarrow}$ is created. An Pauli $X$ gate can be applied to the second spin qubit to transform $\ket{\Psi^+}$ to $\ket{\Phi^+}$.

Multiple pairs of adjacent tree nodes can simultaneously undergo the aforementioned evolution to create Bell states. Then, the entangled spin qubit pairs can be linked by the same procedure. As opposed to having a single photon reflecting off \textit{all the nodes} across each layer to create a GHZ-like state, a process that inevitably suffers from exponentially decaying success probability, the pairwise creation protocol described here is much more efficient.

\subsection{Bell state swap between electron and nuclear spins} \label{sec:spin_swap}

\begin{figure}[h!]
    \centering
    \includegraphics[width=0.5\textwidth]{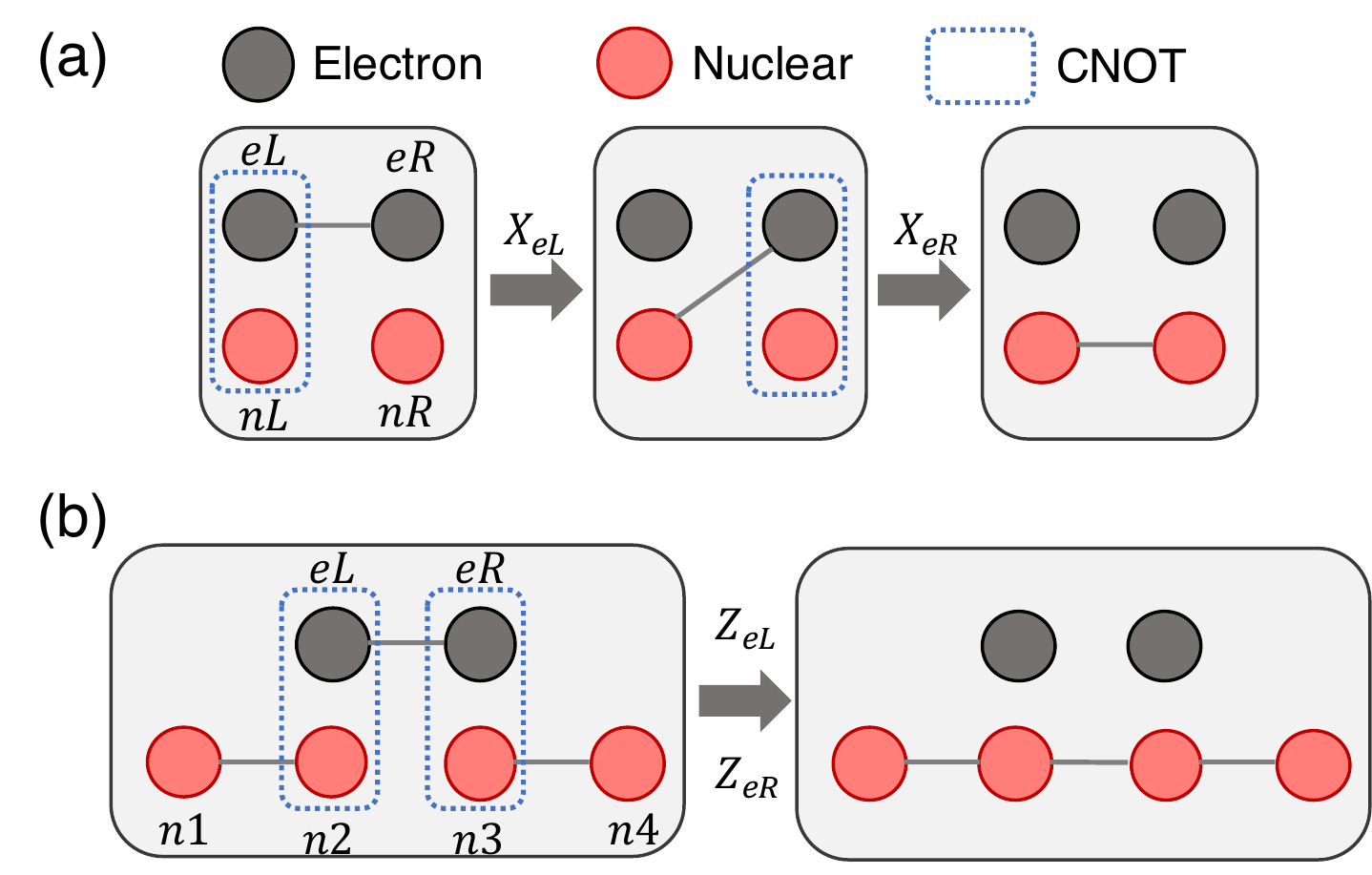}
    \caption{Operations to: (a) swap a Bell state between a pair of entangled electron spins and a pair of nuclear spins; (b) entangle two pairs of Bell states to form a 4-qubit GHZ state in the nuclear spins.}
    \label{fig:supp:Bell_GHZ_states}
\end{figure}

Fig.~\ref{fig:supp:Bell_GHZ_states}(a) shows two electron spins $eL$ and $eR$ entangled in a Bell state: $\ket{00}_e+\ket{11}_e$. Let the nuclear spins initialized in the ground state $\ket{0}_{nL/nR}$. A CNOT operation where $eL$ acts as the control and $nL$ as the target yields an effective GHZ state: $\ket{0}_{nL}\ket{00}_e+\ket{1}_{nL}\ket{11}_e$. Then, an $X$ measurement on $eL$ disentangles the electron spin from the GHZ state, leaving the final state $\ket{\psi}$:
\begin{align}
\ket{\psi} &= \left(\bra{0}\pm\bra{1}\right)_{eL}\left(\ket{0}_{nL}\ket{00}_e+\ket{1}_{nL}\ket{11}_e\right)\\
&= \ket{0}_{nL}\ket{0}_{eR}\pm\ket{1}_{nL}\ket{1}_{eR}
\end{align}

Similarly, a CNOT operation between $eR$ and $nR$ produces $\ket{0}_{nL}\ket{0}_{eR}\ket{0}_{nR}\pm\ket{1}_{nL}\ket{1}_{eR}\ket{1}_{nR}$. A subsequent $X$ measurement on $eR$ then leaves a Bell state between the nuclear spins:
\begin{align}
\ket{\psi} &= \left(\bra{0}\pm\bra{1}\right)_{eR}\left(\ket{0}_{eR}\ket{00}_n\pm\ket{1}_{eR}\ket{11}_n\right)\\
&= \ket{00}_n\pm\ket{11}_n
\end{align}

\subsection{GHZ state creation}

Now, we assume two adjacent pairs of nuclear spins, $\{n1,n2\}$ and $\{n3,n4\}$ are entangled in a Bell state, as shown in Fig.~\ref{fig:supp:Bell_GHZ_states}(b). $n2$ and $n3$'s corresponding electron spins are also entangled in a Bell state via a photon-assisted interaction. We first consider the composite state including $n1,n2,eL,eR$ after a CNOT operation between $n2$ and $eL$, in which $n2$ is the control and $eL$ is the target:
\begin{align}
\ket{\psi} &= \ket{00}_n\ket{00}_e+\ket{00}_n\ket{11}_e\nonumber\\
&\quad +\ket{11}_n\ket{10}_e+\ket{11}_n\ket{01}_e\\
&= \left(\ket{00}_n\ket{0}_{eR}+\ket{11}_n\ket{1}_{eR}\right)\ket{0}_{eL}\nonumber\\
&\quad +\left(\ket{00}_n\ket{1}_{eR}+\ket{11}_n\ket{0}_{eR}\right)\ket{1}_{eL}
\end{align}
where $\ket{ij}_e=\ket{i}_{eL}\ket{j}_{eR}$. A subsequent $Z$ measurement on $eL$ followed by a conditional Pauli transformation on $eR$ yields a GHZ state: $\ket{00}_n\ket{0}_{eR}+\ket{11}\ket{1}_{eR}$.

Then, similarly, a CNOT operation between $n3$ and $eR$ followed by a $Z$ measurement on $eR$ yields the final GHZ state (conditional Pauli transformation on the nuclear spins):
\begin{align}
\ket{\psi} &= \ket{0000}_n+\ket{1111}_n
\end{align}

\subsection{Teleportation}

We present here an example of mapping 2-register addresses $\sum_j \alpha_j\ket{k_{1,j}k_{2,j}}$ onto a 2-level binary tree. Suppose the query addresses compose the superposition state, $\alpha\ket{00}+\beta\ket{01}+\gamma\ket{10}+\delta\ket{11}$, where each register represents the state of the corresponding node at each tree level. We consider the formalism that the atomic state $\ket{0}$ routes the subsequent qubit to the left branch, and $\ket{1}$ to the right. For an instance, the address $\ket{01}$ means the root (level 1) node is in the state $\ket{0}$ and the left node of level 2 is in the state of $\ket{1}$.

\begin{figure}[h!]
    \centering
    \includegraphics[width=0.35\textwidth]{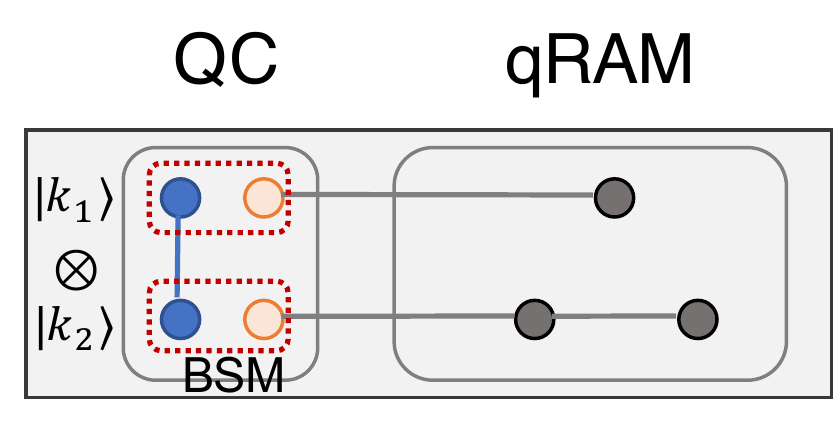}
    \caption{A 2-level qRAM is first entangled with a remote QC. Local BSMs in the QC complete quantum teleportation of the query addresses onto the binary tree. The memory layer is not shown in the schematic for simplicity.}
    \label{fig:supp:teleportation_twolevel}
\end{figure}

Each layer in the qRAM is initialized as a GHZ state, e.g. $\left(\ket{\tilde{0}}+\ket{\tilde{1}}\right)/\sqrt{2}$ where $\ket{\tilde{i}}=\ket{ii...i}$. Importantly, the first register of each GHZ state belongs to an ancillary qubit in the QC, as shown in Fig.~\ref{fig:supp:teleportation_twolevel}.

The un-normalized composite state would then be:
\begin{align}
\ket{\Psi} &= \left(\alpha\ket{00}+\beta\ket{01}+\gamma\ket{10}+\delta\ket{11}\right)\nonumber\\
&\quad\otimes (\ket{00}+\ket{11})_1 (\ket{000}+\ket{111})_2
\end{align}
where the subscripts 1 and 2 denote the layer number.

The state can be re-written as:
\begin{align}
\ket{\Psi}&=\alpha\left[(\ket{\Phi^+}+\ket{\Phi^-})\ket{0}+(\ket{\Psi^+}+\ket{\Psi^-})\ket{1}\right]_1\nonumber\\
&\quad\otimes \left[(\ket{\Phi^+}+\ket{\Phi^-})\ket{00}+(\ket{\Psi^+}+\ket{\Psi^-})\ket{11}\right]_2\\
&+\beta\left[(\ket{\Phi^+}+\ket{\Phi^-})\ket{0}+(\ket{\Psi^+}+\ket{\Psi^-})\ket{1}\right]_1\nonumber\\
&\quad\otimes \left[(\ket{\Psi^+}-\ket{\Psi^-})\ket{00}+(\ket{\Phi^+}-\ket{\Phi^-})\ket{11}\right]_2\\
&+\gamma\left[(\ket{\Psi^+}-\ket{\Psi^-})\ket{0}+(\ket{\Phi^+}-\ket{\Phi^-})\ket{1}\right]_1\nonumber\\
&\quad\otimes \left[(\ket{\Phi^+}+\ket{\Phi^-})\ket{00}+(\ket{\Psi^+}+\ket{\Psi^-})\ket{11}\right]_2\\
&+\delta\left[(\ket{\Psi^+}-\ket{\Psi^-})\ket{0}+(\ket{\Phi^+}-\ket{\Phi^-})\ket{1}\right]_1\nonumber\\
&\quad\otimes \left[(\ket{\Psi^+}-\ket{\Psi^-})\ket{00}+(\ket{\Phi^+}-\ket{\Phi^-})\ket{11}\right]_2\\
&= \ket{\Phi^+}_1\ket{\Phi^+}_2 \left( \alpha\ket{0}_1\ket{00}_2+\beta\ket{0}_1\ket{11}_2\right.\nonumber\\
&\quad\quad +\left.\gamma\ket{1}_1\ket{00}_2+\delta\ket{1}_1\ket{11}_2\right)+...
\end{align}
Bell state measurements for each layer would then project the composite state into one of the 16 possible combinations. Followed by conditional Pauli transformations, the query addresses are finally teleported onto the binary tree.

\subsection{Efficiency simulations} \label{sec:efficiency_simulations}

The teleportation scheme includes 4 steps: (1) initializing the entanglement links, (2) teleporting the addresses to the qRAM, (3) querying, and (4) teleporting the addresses back to the QC.  We perform event-based simulations to estimate the time of completing all four steps.

In step (1), all the nodes except the leftmost node within each qRAM layer are entangled to form a GHZ state. During its creation process, pairs of the nearest neighbors are first entangled by heralding, with success probability $p_{\text{ep}}=\eta_{\text{path}}\eta_s^2\eta_{\text{det}}$ (see App.~\ref{sec:efficiency}). If the entanglement attempt fails, the spins undergo re-initialization for $\treset=5~\mu$s. If it succeeds, the electron spins (broker qubits) are swapped with their respective nuclear spins (memory qubits), an operation which we assume to take $t_{e\rightarrow n}=16~\mu$s. Then, the unlinked neighbors are subsequently entangled in the same fashion. To reduce computational costs, we assume the rate is limited by the largest layer and only simulate its GHZ state creation process.

Simultaneously in step (2), we attempt to generate entanglement between the QC's broker qubit and the qRAM's leftmost node for each layer. Once the entanglement link is generated, the electron and nuclear spins are again swapped. In simulation, we take the maximum between the time to generate a GHZ state and the time to produce QC-qRAM Bell state. Once both states are constructed, the leftmost node is entangled with the GHZ state composed of the remaining nodes within the same layer. Then, a local BSM is made between the address register and the QC ancillary qubits. To fairly compare the teleportation scheme's efficiency with the GLM scheme, we neglect the physical distance between the QC and the qRAM in Fig.~\ref{fig:teleportation_efficiency}.

In step (3), a bus photon arrives at the root node of the binary tree and is routed to the memory layer with the query success probability $p_i$ for an $i$-level qRAM. Finally, in step (4), a QC-qRAM Bell state is constructed again for each layer with probability $p_{\text{ep}}$, followed by local BSMs on the leftmost nodes in the qRAM.

In Fig.~\ref{fig:teleportation_efficiency}, the simulation data are plotted along with their analytical fits. Recall that the GHZ states are produced by linking multiples of Bell pairs. If each Bell pair creation succeeds with probability $p$, it would take a geometric mean of $1/p$ attempts. In the case of $p=1$, the GHZ state creation process would merely be a two-step process. For example, for a layer with 4 nodes, nodes 1 and 2 as well as nodes 3 and 4 are entangled in the first time step. Then, nodes 2 and 3 are entangled to complete the GHZ state creation. However, with a non-unity $p$, the GHZ state creation is ultimately limited by the pair that fails the most number of times. In other words, the rate is mainly determined by the \textit{outlier}. We fit the guessed model $f(N)=aN^{-b}$ multiplied with the analytical rate (based on geometric mean) to the simulation data, where $N$ is the number of nodes within the largest layer. The coefficients $a,b$ capture the outlier's scaling with the circuit depth. Their fitted values averaged over the considered $\kappa_{\text{wg}}/\kappa$ ratios are summarized in Table~\ref{tab:parameters}.

\begin{table}[h!]
\caption{\label{tab:parameters} Parameter values.}
\begin{ruledtabular}
\begin{tabular}{cc}
$\omega_c$ & 406.774~THz\\
\hline
$\kappa$ & 20.34~GHz~\cite{Bhaskar_2019}\\
\hline
$\gamma$ & 94~MHz\\
\hline
$\treset$ & 5~$\mu$s\\
\hline
$\eta_{\text{str}}$ & 2.7~dB~\cite{Desiatov_2019}\\
\hline
$\eta_{\text{bend}}$ & 9.3~dB~\cite{Desiatov_2019}\\
\hline
$\eta_{\text{det}}$ & 1.3~dB~\cite{Joshi_2020}\\
\hline
$R_{\text{resonator}}$ & 50~$\mu$m~\cite{Zhang_2017}\\
\hline
$\neff$ & 2.2645\\
\hline
$n_{g,\text{PIC}}$ & 2.3862\\
\hline
$n_{g,\text{dmd}}$ & 2.4513\\
\hline
$t_{e\rightarrow n}$ & 16~$\mu$s\\
\hline
$t_{n\rightarrow e}$ & 30~ns\\
\hline
$a$ & 1.7094\\
\hline
$b$ & 0.79386\\
\end{tabular}
\end{ruledtabular}
\end{table}

\begin{figure}[h!]
    \centering
    \includegraphics[width=0.5\textwidth]{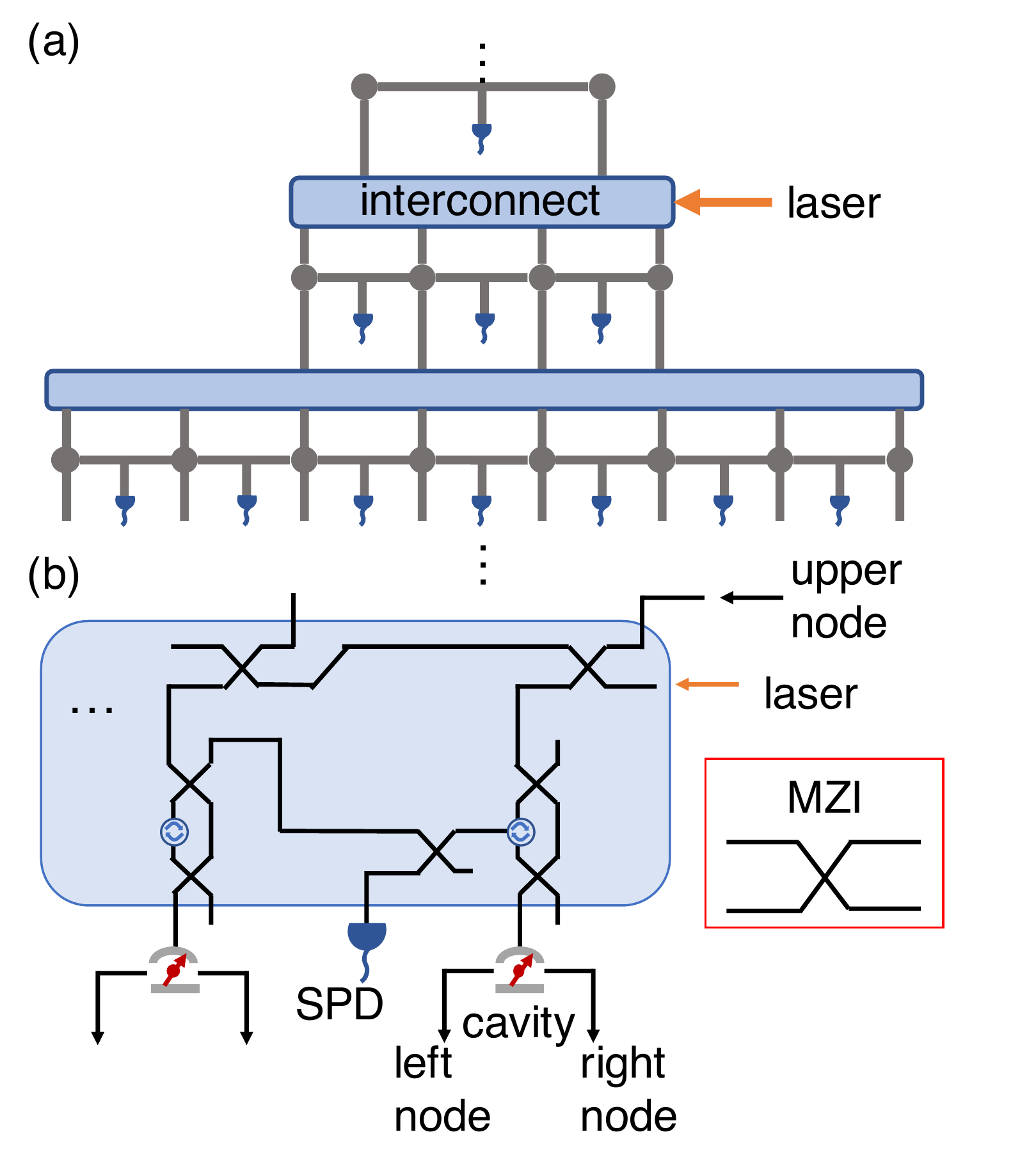}
    \caption{Proposed PIC architecture for the teleportation scheme. (a) The qRAM binary tree contains interspersed interconnect layers that enable intra-layer connectivity. (b) Within each interconnect layer, a network of MZIs is classically controlled to direct the single photons to either the subsequent cavity or the detection system for heralding during GHZ state creation. It is then switched to a transparent state during the data retrieval step.}
    \label{fig:supp:PIC_interconnect}
\end{figure}

\subsection{PIC interconnect} \label{sec:PIC_interconnect}

In contrast with the GLM scheme, the teleportation scheme requires greater connectivity in the qRAM. Each node is not only connected to two children nodes in the next layer, but also to the rest of the nodes \textit{in the same layer}. Here, we detail its PIC construct. Importantly, as shown in Fig.~\ref{fig:supp:PIC_interconnect}(a), the architecture requires interconnect layers interspersed between the binary tree layers. Additionally, a photon detection system resides between each neighboring pair. Assuming the single photons are propagating in one direction, i.e. incoming from the right of each layer, the detector would register photons after they interact with the cavities to its right.

Within each interconnect layer, MZI switches are classically controlled to enable routing the single photons to individual cavities. The cavity depicted in Fig.~\ref{fig:supp:PIC_interconnect}(b) is the same construct shown in Fig.~\ref{fig:supp:implementation}. To entangle two neighboring nodes, each tree node first operates in the setting mode. A single photon reflects off the first cavity and is directed to the second cavity via a circulator. After entering the second node through the MZI and reflecting off the cavity, the photon is again routed to an MZI switch via a circulator. Except now, the switch directs the photon to the detector for heralding a Bell state creation. We stress here that the interconnect layer enables beyond nearest neighbor connection. Therefore, given prior knowledge of the query addresses, the architecture provides the ability to only entangle the necessary nodes and reduces state infidelity.

\begin{figure}[h!]
    \centering
    \includegraphics[width=0.5\textwidth]{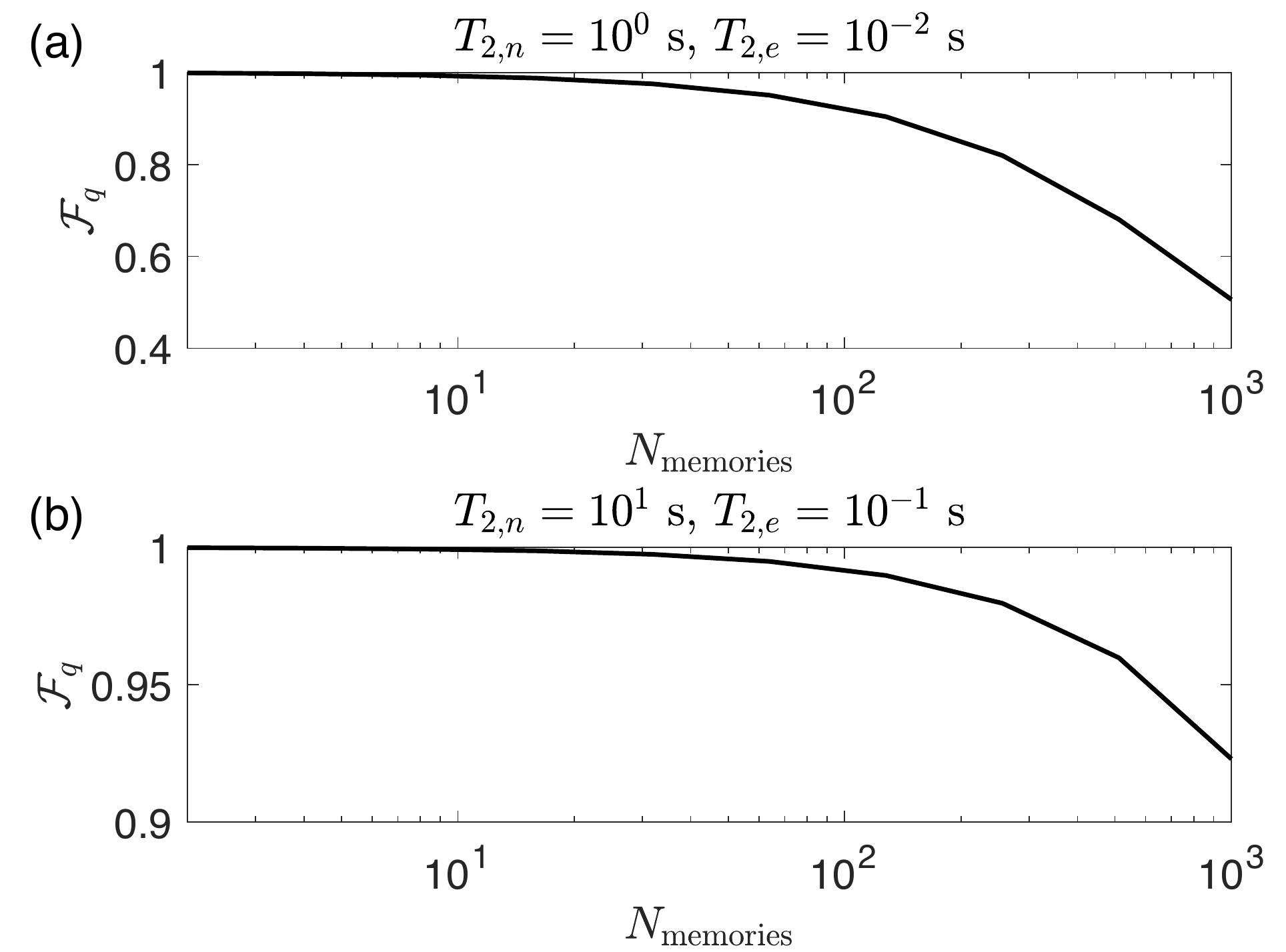}
    \caption{Query fidelity as a function of qRAM size. The nuclear and electron spin coherence times are respectively assumed to be: (a) $T_{2,n}=10^0~$s and $T_{2,e}=10^{-2}~$s, (b) $T_{2,n}=10^1~$s and $T_{2,e}=10^{-1}~$s.}
    \label{fig:supp:decoherence_only}
\end{figure}

After the addresses are teleported from the QC to the qRAM, the cavity nodes are changed to the routing mode to direct the bus qubit to the memory layer at the bottom of the binary tree. In this step, the interconnect layer is essentially transparent by having the photon bypassing the circulators.

\subsection{Infidelity from decoherence} \label{sec:decoherence}

We define the query fidelity as the fidelity of the prepared tree state. To calculate infidelity caused by decoherence for each layer, we take the decoherence rate $\gamma_d$ to be proportional to $Nt_{e\rightarrow n}/T_{2,n/e}$, where $N$ is the number of nodes and $t_{e\rightarrow n}$ is the approximate entanglement time. We assume perfect single qubit rotations, readout, and setting fidelity by optimally balancing losses (see App.~\ref{sec:fidelity}). We only consider the effect of decoherence caused by continuous dephasing and neglect other physical errors such as imperfect nuclear-electron spin interaction. In Fig.~\ref{fig:supp:decoherence_only}(a), at $N_{\text{memories}}=10^3$, the query fidelity drops to near $\mathcal{F}_q=0.5$, suggesting the prepared tree is no better than a maximally mixed state. Improvements on the nuclear and electron spin coherence times by an order-of-magnitude can increase the fidelity to $\mathcal{F}_q>0.9$, as shown in Fig.~\ref{fig:supp:decoherence_only}(b).

\bibliography{references_qram}

\begin{thebibliography}{38}%
\makeatletter
\providecommand \@ifxundefined [1]{%
 \@ifx{#1\undefined}
}%
\providecommand \@ifnum [1]{%
 \ifnum #1\expandafter \@firstoftwo
 \else \expandafter \@secondoftwo
 \fi
}%
\providecommand \@ifx [1]{%
 \ifx #1\expandafter \@firstoftwo
 \else \expandafter \@secondoftwo
 \fi
}%
\providecommand \natexlab [1]{#1}%
\providecommand \enquote  [1]{``#1''}%
\providecommand \bibnamefont  [1]{#1}%
\providecommand \bibfnamefont [1]{#1}%
\providecommand \citenamefont [1]{#1}%
\providecommand \href@noop [0]{\@secondoftwo}%
\providecommand \href [0]{\begingroup \@sanitize@url \@href}%
\providecommand \@href[1]{\@@startlink{#1}\@@href}%
\providecommand \@@href[1]{\endgroup#1\@@endlink}%
\providecommand \@sanitize@url [0]{\catcode `\\12\catcode `\$12\catcode
  `\&12\catcode `\#12\catcode `\^12\catcode `\_12\catcode `\%12\relax}%
\providecommand \@@startlink[1]{}%
\providecommand \@@endlink[0]{}%
\providecommand \url  [0]{\begingroup\@sanitize@url \@url }%
\providecommand \@url [1]{\endgroup\@href {#1}{\urlprefix }}%
\providecommand \urlprefix  [0]{URL }%
\providecommand \Eprint [0]{\href }%
\providecommand \doibase [0]{http://dx.doi.org/}%
\providecommand \selectlanguage [0]{\@gobble}%
\providecommand \bibinfo  [0]{\@secondoftwo}%
\providecommand \bibfield  [0]{\@secondoftwo}%
\providecommand \translation [1]{[#1]}%
\providecommand \BibitemOpen [0]{}%
\providecommand \bibitemStop [0]{}%
\providecommand \bibitemNoStop [0]{.\EOS\space}%
\providecommand \EOS [0]{\spacefactor3000\relax}%
\providecommand \BibitemShut  [1]{\csname bibitem#1\endcsname}%
\let\auto@bib@innerbib\@empty
\bibitem [{\citenamefont {Giovannetti}\ \emph
  {et~al.}(2008{\natexlab{a}})\citenamefont {Giovannetti}, \citenamefont
  {Lloyd},\ and\ \citenamefont {Maccone}}]{Giovannetti_2008_PRL}%
  \BibitemOpen
  \bibfield  {author} {\bibinfo {author} {\bibfnamefont {V.}~\bibnamefont
  {Giovannetti}}, \bibinfo {author} {\bibfnamefont {S.}~\bibnamefont {Lloyd}},
  \ and\ \bibinfo {author} {\bibfnamefont {L.}~\bibnamefont {Maccone}},\
  }\href@noop {} {\bibfield  {journal} {\bibinfo  {journal} {Phys. Rev. Lett.}\
  }\textbf {\bibinfo {volume} {100}},\ \bibinfo {pages} {160501} (\bibinfo
  {year} {2008}{\natexlab{a}})}\BibitemShut {NoStop}%
\bibitem [{\citenamefont {Giovannetti}\ \emph
  {et~al.}(2008{\natexlab{b}})\citenamefont {Giovannetti}, \citenamefont
  {Lloyd},\ and\ \citenamefont {Maccone}}]{Giovannetti_2008_PRA}%
  \BibitemOpen
  \bibfield  {author} {\bibinfo {author} {\bibfnamefont {V.}~\bibnamefont
  {Giovannetti}}, \bibinfo {author} {\bibfnamefont {S.}~\bibnamefont {Lloyd}},
  \ and\ \bibinfo {author} {\bibfnamefont {L.}~\bibnamefont {Maccone}},\
  }\href@noop {} {\bibfield  {journal} {\bibinfo  {journal} {Phys. Rev. A}\
  }\textbf {\bibinfo {volume} {78}},\ \bibinfo {pages} {052310} (\bibinfo
  {year} {2008}{\natexlab{b}})}\BibitemShut {NoStop}%
\bibitem [{\citenamefont {Biamonte}\ \emph {et~al.}(2017)\citenamefont
  {Biamonte}, \citenamefont {Wittek}, \citenamefont {Pancotti}, \citenamefont
  {Rebentrost}, \citenamefont {Wiebe},\ and\ \citenamefont
  {Lloyd}}]{Biamonte_2017}%
  \BibitemOpen
  \bibfield  {author} {\bibinfo {author} {\bibfnamefont {J.}~\bibnamefont
  {Biamonte}}, \bibinfo {author} {\bibfnamefont {P.}~\bibnamefont {Wittek}},
  \bibinfo {author} {\bibfnamefont {N.}~\bibnamefont {Pancotti}}, \bibinfo
  {author} {\bibfnamefont {P.}~\bibnamefont {Rebentrost}}, \bibinfo {author}
  {\bibfnamefont {N.}~\bibnamefont {Wiebe}}, \ and\ \bibinfo {author}
  {\bibfnamefont {S.}~\bibnamefont {Lloyd}},\ }\href@noop {} {\bibfield
  {journal} {\bibinfo  {journal} {Nature}\ }\textbf {\bibinfo {volume} {549}},\
  \bibinfo {pages} {195–202} (\bibinfo {year} {2017})}\BibitemShut {NoStop}%
\bibitem [{\citenamefont {Harrow}\ \emph {et~al.}(2009)\citenamefont {Harrow},
  \citenamefont {Hassidim},\ and\ \citenamefont {Lloyd}}]{Harrow_2009}%
  \BibitemOpen
  \bibfield  {author} {\bibinfo {author} {\bibfnamefont {A.~W.}\ \bibnamefont
  {Harrow}}, \bibinfo {author} {\bibfnamefont {A.}~\bibnamefont {Hassidim}}, \
  and\ \bibinfo {author} {\bibfnamefont {S.}~\bibnamefont {Lloyd}},\
  }\href@noop {} {\bibfield  {journal} {\bibinfo  {journal} {Phys. Rev. Lett.}\
  }\textbf {\bibinfo {volume} {103}},\ \bibinfo {pages} {150502} (\bibinfo
  {year} {2009})}\BibitemShut {NoStop}%
\bibitem [{\citenamefont {Kiani}\ \emph {et~al.}(2020)\citenamefont {Kiani},
  \citenamefont {Villanyi},\ and\ \citenamefont {Lloyd}}]{Kiani_2020}%
  \BibitemOpen
  \bibfield  {author} {\bibinfo {author} {\bibfnamefont {B.~T.}\ \bibnamefont
  {Kiani}}, \bibinfo {author} {\bibfnamefont {A.}~\bibnamefont {Villanyi}}, \
  and\ \bibinfo {author} {\bibfnamefont {S.}~\bibnamefont {Lloyd}},\
  }\href@noop {} {\bibfield  {journal} {\bibinfo  {journal} {ArXiv}\ ,\
  \bibinfo {pages} {2004.02036}} (\bibinfo {year} {2020})}\BibitemShut
  {NoStop}%
\bibitem [{\citenamefont {Grover}(1996)}]{Grover_1996}%
  \BibitemOpen
  \bibfield  {author} {\bibinfo {author} {\bibfnamefont {L.}~\bibnamefont
  {Grover}},\ }\bibinfo {organization} {Proceedings, 28th Annual ACM Symposium
  on the Theory of Computing}\ (\bibinfo  {publisher} {ACM Press},\ \bibinfo
  {year} {1996})\BibitemShut {NoStop}%
\bibitem [{\citenamefont {Hong}\ \emph {et~al.}(2012)\citenamefont {Hong},
  \citenamefont {Xiang}, \citenamefont {Zhu}, \citenamefont {zhen Jiang},\ and\
  \citenamefont {neng Wu}}]{Hong_2012}%
  \BibitemOpen
  \bibfield  {author} {\bibinfo {author} {\bibfnamefont {F.-Y.}\ \bibnamefont
  {Hong}}, \bibinfo {author} {\bibfnamefont {Y.}~\bibnamefont {Xiang}},
  \bibinfo {author} {\bibfnamefont {Z.-Y.}\ \bibnamefont {Zhu}}, \bibinfo
  {author} {\bibfnamefont {L.}~\bibnamefont {zhen Jiang}}, \ and\ \bibinfo
  {author} {\bibfnamefont {L.}~\bibnamefont {neng Wu}},\ }\href@noop {}
  {\bibfield  {journal} {\bibinfo  {journal} {Phys. Rev. A}\ }\textbf {\bibinfo
  {volume} {86}},\ \bibinfo {pages} {010306} (\bibinfo {year}
  {2012})}\BibitemShut {NoStop}%
\bibitem [{\citenamefont {Moiseev}\ and\ \citenamefont
  {Moiseev}(2016)}]{Moiseev_2016}%
  \BibitemOpen
  \bibfield  {author} {\bibinfo {author} {\bibfnamefont {E.~S.}\ \bibnamefont
  {Moiseev}}\ and\ \bibinfo {author} {\bibfnamefont {S.~A.}\ \bibnamefont
  {Moiseev}},\ }\href@noop {} {\bibfield  {journal} {\bibinfo  {journal} {J.
  Mod. Opt.}\ }\textbf {\bibinfo {volume} {63}},\ \bibinfo {pages}
  {2081–2092} (\bibinfo {year} {2016})}\BibitemShut {NoStop}%
\bibitem [{\citenamefont {Hann}\ \emph {et~al.}(2019)\citenamefont {Hann},
  \citenamefont {Zou}, \citenamefont {Zhang}, \citenamefont {Chu},
  \citenamefont {Schoelkopf}, \citenamefont {Girvin},\ and\ \citenamefont
  {Jiang}}]{Hann_2019}%
  \BibitemOpen
  \bibfield  {author} {\bibinfo {author} {\bibfnamefont {C.~T.}\ \bibnamefont
  {Hann}}, \bibinfo {author} {\bibfnamefont {C.-L.}\ \bibnamefont {Zou}},
  \bibinfo {author} {\bibfnamefont {Y.}~\bibnamefont {Zhang}}, \bibinfo
  {author} {\bibfnamefont {Y.}~\bibnamefont {Chu}}, \bibinfo {author}
  {\bibfnamefont {R.~J.}\ \bibnamefont {Schoelkopf}}, \bibinfo {author}
  {\bibfnamefont {S.~M.}\ \bibnamefont {Girvin}}, \ and\ \bibinfo {author}
  {\bibfnamefont {L.}~\bibnamefont {Jiang}},\ }\href@noop {} {\bibfield
  {journal} {\bibinfo  {journal} {Phys. Rev. Lett.}\ }\textbf {\bibinfo
  {volume} {123}},\ \bibinfo {pages} {250501} (\bibinfo {year}
  {2019})}\BibitemShut {NoStop}%
\bibitem [{\citenamefont {Lu}\ \emph {et~al.}(2018)\citenamefont {Lu},
  \citenamefont {Lukens}, \citenamefont {Peters}, \citenamefont {Odele},
  \citenamefont {Leaird}, \citenamefont {Weiner},\ and\ \citenamefont
  {Lougovski}}]{Lu_2018}%
  \BibitemOpen
  \bibfield  {author} {\bibinfo {author} {\bibfnamefont {H.-H.}\ \bibnamefont
  {Lu}}, \bibinfo {author} {\bibfnamefont {J.~M.}\ \bibnamefont {Lukens}},
  \bibinfo {author} {\bibfnamefont {N.~A.}\ \bibnamefont {Peters}}, \bibinfo
  {author} {\bibfnamefont {O.~D.}\ \bibnamefont {Odele}}, \bibinfo {author}
  {\bibfnamefont {D.~E.}\ \bibnamefont {Leaird}}, \bibinfo {author}
  {\bibfnamefont {A.~M.}\ \bibnamefont {Weiner}}, \ and\ \bibinfo {author}
  {\bibfnamefont {P.}~\bibnamefont {Lougovski}},\ }\href@noop {} {\bibfield
  {journal} {\bibinfo  {journal} {Phys. Rev. Lett.}\ }\textbf {\bibinfo
  {volume} {120}},\ \bibinfo {pages} {030502} (\bibinfo {year}
  {2018})}\BibitemShut {NoStop}%
\bibitem [{\citenamefont {Joshi}\ \emph {et~al.}(2020)\citenamefont {Joshi},
  \citenamefont {Farsi}, \citenamefont {Dutt}, \citenamefont {Kim},
  \citenamefont {Ji}, \citenamefont {Zhao}, \citenamefont {Bishop},
  \citenamefont {Lipson},\ and\ \citenamefont {Gaeta}}]{Joshi_2020}%
  \BibitemOpen
  \bibfield  {author} {\bibinfo {author} {\bibfnamefont {C.}~\bibnamefont
  {Joshi}}, \bibinfo {author} {\bibfnamefont {A.}~\bibnamefont {Farsi}},
  \bibinfo {author} {\bibfnamefont {A.}~\bibnamefont {Dutt}}, \bibinfo {author}
  {\bibfnamefont {B.~Y.}\ \bibnamefont {Kim}}, \bibinfo {author} {\bibfnamefont
  {X.}~\bibnamefont {Ji}}, \bibinfo {author} {\bibfnamefont {Y.}~\bibnamefont
  {Zhao}}, \bibinfo {author} {\bibfnamefont {A.~M.}\ \bibnamefont {Bishop}},
  \bibinfo {author} {\bibfnamefont {M.}~\bibnamefont {Lipson}}, \ and\ \bibinfo
  {author} {\bibfnamefont {A.~L.}\ \bibnamefont {Gaeta}},\ }\href@noop {}
  {\bibfield  {journal} {\bibinfo  {journal} {Phys. Rev. Lett.}\ }\textbf
  {\bibinfo {volume} {124}},\ \bibinfo {pages} {143601} (\bibinfo {year}
  {2020})}\BibitemShut {NoStop}%
\bibitem [{\citenamefont {Lu}\ \emph {et~al.}(2020)\citenamefont {Lu},
  \citenamefont {Simmerman}, \citenamefont {Lougovski}, \citenamefont
  {Weiner},\ and\ \citenamefont {Lukens}}]{Lu_2020}%
  \BibitemOpen
  \bibfield  {author} {\bibinfo {author} {\bibfnamefont {H.-H.}\ \bibnamefont
  {Lu}}, \bibinfo {author} {\bibfnamefont {E.~M.}\ \bibnamefont {Simmerman}},
  \bibinfo {author} {\bibfnamefont {P.}~\bibnamefont {Lougovski}}, \bibinfo
  {author} {\bibfnamefont {A.~M.}\ \bibnamefont {Weiner}}, \ and\ \bibinfo
  {author} {\bibfnamefont {J.~M.}\ \bibnamefont {Lukens}},\ }\href@noop {}
  {\bibfield  {journal} {\bibinfo  {journal} {ArXiv}\ ,\ \bibinfo {pages}
  {2008.07444}} (\bibinfo {year} {2020})}\BibitemShut {NoStop}%
\bibitem [{\citenamefont {Nguyen}\ \emph {et~al.}(2019)\citenamefont {Nguyen}
  \emph {et~al.}}]{Nguyen_2019_PRB}%
  \BibitemOpen
  \bibfield  {author} {\bibinfo {author} {\bibfnamefont {C.~T.}\ \bibnamefont
  {Nguyen}} \emph {et~al.},\ }\href@noop {} {\bibfield  {journal} {\bibinfo
  {journal} {Phys. Rev. B}\ }\textbf {\bibinfo {volume} {100}},\ \bibinfo
  {pages} {165428} (\bibinfo {year} {2019})}\BibitemShut {NoStop}%
\bibitem [{\citenamefont {Bhaskar}\ \emph {et~al.}(2020)\citenamefont {Bhaskar}
  \emph {et~al.}}]{Bhaskar_2019}%
  \BibitemOpen
  \bibfield  {author} {\bibinfo {author} {\bibfnamefont {M.~K.}\ \bibnamefont
  {Bhaskar}} \emph {et~al.},\ }\href@noop {} {\bibfield  {journal} {\bibinfo
  {journal} {Nature}\ }\textbf {\bibinfo {volume} {580}},\ \bibinfo {pages}
  {60–64} (\bibinfo {year} {2020})}\BibitemShut {NoStop}%
\bibitem [{\citenamefont {Harris}\ \emph {et~al.}(2017)\citenamefont {Harris},
  \citenamefont {Steinbrecher}, \citenamefont {Prabhu}, \citenamefont {Lahini},
  \citenamefont {Mower}, \citenamefont {Bunandar}, \citenamefont {Chen},
  \citenamefont {Wong}, \citenamefont {Baehr-Jones}, \citenamefont {Hochberg},
  \citenamefont {Lloyd},\ and\ \citenamefont {Englund}}]{Harris_2017}%
  \BibitemOpen
  \bibfield  {author} {\bibinfo {author} {\bibfnamefont {N.~C.}\ \bibnamefont
  {Harris}}, \bibinfo {author} {\bibfnamefont {G.~R.}\ \bibnamefont
  {Steinbrecher}}, \bibinfo {author} {\bibfnamefont {M.}~\bibnamefont
  {Prabhu}}, \bibinfo {author} {\bibfnamefont {Y.}~\bibnamefont {Lahini}},
  \bibinfo {author} {\bibfnamefont {J.}~\bibnamefont {Mower}}, \bibinfo
  {author} {\bibfnamefont {D.}~\bibnamefont {Bunandar}}, \bibinfo {author}
  {\bibfnamefont {C.}~\bibnamefont {Chen}}, \bibinfo {author} {\bibfnamefont
  {F.~N.~C.}\ \bibnamefont {Wong}}, \bibinfo {author} {\bibfnamefont
  {T.}~\bibnamefont {Baehr-Jones}}, \bibinfo {author} {\bibfnamefont
  {M.}~\bibnamefont {Hochberg}}, \bibinfo {author} {\bibfnamefont
  {S.}~\bibnamefont {Lloyd}}, \ and\ \bibinfo {author} {\bibfnamefont
  {D.}~\bibnamefont {Englund}},\ }\href@noop {} {\bibfield  {journal} {\bibinfo
   {journal} {Nature Photon.}\ }\textbf {\bibinfo {volume} {11}},\ \bibinfo
  {pages} {447–452} (\bibinfo {year} {2017})}\BibitemShut {NoStop}%
\bibitem [{\citenamefont {Duan}\ and\ \citenamefont
  {Kimble}(2004)}]{Duan_2004}%
  \BibitemOpen
  \bibfield  {author} {\bibinfo {author} {\bibfnamefont {L.-M.}\ \bibnamefont
  {Duan}}\ and\ \bibinfo {author} {\bibfnamefont {H.~J.}\ \bibnamefont
  {Kimble}},\ }\href@noop {} {\bibfield  {journal} {\bibinfo  {journal} {Phys.
  Rev. Lett.}\ }\textbf {\bibinfo {volume} {92}},\ \bibinfo {pages} {127902}
  (\bibinfo {year} {2004})}\BibitemShut {NoStop}%
\bibitem [{\citenamefont {Nielsen}\ and\ \citenamefont
  {Chuang}(2010)}]{Chuang_2010}%
  \BibitemOpen
  \bibfield  {author} {\bibinfo {author} {\bibfnamefont {M.~A.}\ \bibnamefont
  {Nielsen}}\ and\ \bibinfo {author} {\bibfnamefont {I.~L.}\ \bibnamefont
  {Chuang}},\ }\href@noop {} {\emph {\bibinfo {title} {Quantum Computation and
  Quantum Information}}},\ \bibinfo {edition} {10th}\ ed.\ (\bibinfo
  {publisher} {Cambridge University Press},\ \bibinfo {year}
  {2010})\BibitemShut {NoStop}%
\bibitem [{\citenamefont {Bowdrey}\ \emph {et~al.}(2002)\citenamefont
  {Bowdrey}, \citenamefont {Oi}, \citenamefont {Short}, \citenamefont
  {Banaszek},\ and\ \citenamefont {Jones}}]{Bowdrey_2002}%
  \BibitemOpen
  \bibfield  {author} {\bibinfo {author} {\bibfnamefont {M.~D.}\ \bibnamefont
  {Bowdrey}}, \bibinfo {author} {\bibfnamefont {D.~K.}\ \bibnamefont {Oi}},
  \bibinfo {author} {\bibfnamefont {A.~J.}\ \bibnamefont {Short}}, \bibinfo
  {author} {\bibfnamefont {K.}~\bibnamefont {Banaszek}}, \ and\ \bibinfo
  {author} {\bibfnamefont {J.~A.}\ \bibnamefont {Jones}},\ }\href@noop {}
  {\bibfield  {journal} {\bibinfo  {journal} {Phys. Lett. A}\ }\textbf
  {\bibinfo {volume} {294}},\ \bibinfo {pages} {258–260} (\bibinfo {year}
  {2002})}\BibitemShut {NoStop}%
\bibitem [{\citenamefont {Hu}\ \emph {et~al.}(2008)\citenamefont {Hu},
  \citenamefont {Young}, \citenamefont {O’Brien}, \citenamefont {Munro},\
  and\ \citenamefont {Rarity}}]{Hu_2008}%
  \BibitemOpen
  \bibfield  {author} {\bibinfo {author} {\bibfnamefont {C.~Y.}\ \bibnamefont
  {Hu}}, \bibinfo {author} {\bibfnamefont {A.}~\bibnamefont {Young}}, \bibinfo
  {author} {\bibfnamefont {J.~L.}\ \bibnamefont {O’Brien}}, \bibinfo {author}
  {\bibfnamefont {W.~J.}\ \bibnamefont {Munro}}, \ and\ \bibinfo {author}
  {\bibfnamefont {J.~G.}\ \bibnamefont {Rarity}},\ }\href@noop {} {\bibfield
  {journal} {\bibinfo  {journal} {Phys. Rev. B}\ }\textbf {\bibinfo {volume}
  {78}},\ \bibinfo {pages} {085307} (\bibinfo {year} {2008})}\BibitemShut
  {NoStop}%
\bibitem [{\citenamefont {Tiecke}\ \emph {et~al.}(2014)\citenamefont {Tiecke}
  \emph {et~al.}}]{Tiecke_2014}%
  \BibitemOpen
  \bibfield  {author} {\bibinfo {author} {\bibfnamefont {T.~G.}\ \bibnamefont
  {Tiecke}} \emph {et~al.},\ }\href@noop {} {\bibfield  {journal} {\bibinfo
  {journal} {Nature}\ }\textbf {\bibinfo {volume} {508}},\ \bibinfo {pages}
  {241–244} (\bibinfo {year} {2014})}\BibitemShut {NoStop}%
\bibitem [{\citenamefont {Rong}\ \emph {et~al.}(2015)\citenamefont {Rong},
  \citenamefont {Geng}, \citenamefont {Shi}, \citenamefont {Liu}, \citenamefont
  {Xu}, \citenamefont {Ma}, \citenamefont {Kong}, \citenamefont {Jiang},
  \citenamefont {Wu},\ and\ \citenamefont {Du}}]{Rong_2015}%
  \BibitemOpen
  \bibfield  {author} {\bibinfo {author} {\bibfnamefont {X.}~\bibnamefont
  {Rong}}, \bibinfo {author} {\bibfnamefont {J.}~\bibnamefont {Geng}}, \bibinfo
  {author} {\bibfnamefont {F.}~\bibnamefont {Shi}}, \bibinfo {author}
  {\bibfnamefont {Y.}~\bibnamefont {Liu}}, \bibinfo {author} {\bibfnamefont
  {K.}~\bibnamefont {Xu}}, \bibinfo {author} {\bibfnamefont {W.}~\bibnamefont
  {Ma}}, \bibinfo {author} {\bibfnamefont {F.}~\bibnamefont {Kong}}, \bibinfo
  {author} {\bibfnamefont {Z.}~\bibnamefont {Jiang}}, \bibinfo {author}
  {\bibfnamefont {Y.}~\bibnamefont {Wu}}, \ and\ \bibinfo {author}
  {\bibfnamefont {J.}~\bibnamefont {Du}},\ }\href@noop {} {\bibfield  {journal}
  {\bibinfo  {journal} {Nat. Commun.}\ }\textbf {\bibinfo {volume} {6}},\
  \bibinfo {pages} {8748} (\bibinfo {year} {2015})}\BibitemShut {NoStop}%
\bibitem [{\citenamefont {Kalb}\ \emph {et~al.}(2017)\citenamefont {Kalb},
  \citenamefont {Reiserer}, \citenamefont {Humphreys}, \citenamefont
  {Bakermans}, \citenamefont {Kamerling}, \citenamefont {Nickerson},
  \citenamefont {Benjamin}, \citenamefont {Twitchen}, \citenamefont
  {M.Markham},\ and\ \citenamefont {Hanson}}]{Kalb_2017}%
  \BibitemOpen
  \bibfield  {author} {\bibinfo {author} {\bibfnamefont {N.}~\bibnamefont
  {Kalb}}, \bibinfo {author} {\bibfnamefont {A.~A.}\ \bibnamefont {Reiserer}},
  \bibinfo {author} {\bibfnamefont {P.~C.}\ \bibnamefont {Humphreys}}, \bibinfo
  {author} {\bibfnamefont {J.~J.~W.}\ \bibnamefont {Bakermans}}, \bibinfo
  {author} {\bibfnamefont {S.~J.}\ \bibnamefont {Kamerling}}, \bibinfo {author}
  {\bibfnamefont {N.~H.}\ \bibnamefont {Nickerson}}, \bibinfo {author}
  {\bibfnamefont {S.~C.}\ \bibnamefont {Benjamin}}, \bibinfo {author}
  {\bibfnamefont {D.~J.}\ \bibnamefont {Twitchen}}, \bibinfo {author}
  {\bibnamefont {M.Markham}}, \ and\ \bibinfo {author} {\bibfnamefont
  {R.}~\bibnamefont {Hanson}},\ }\href@noop {} {\bibfield  {journal} {\bibinfo
  {journal} {Science}\ }\textbf {\bibinfo {volume} {356}},\ \bibinfo {pages}
  {928–932} (\bibinfo {year} {2017})}\BibitemShut {NoStop}%
\bibitem [{\citenamefont {Sukachev}\ \emph {et~al.}(2017)\citenamefont
  {Sukachev}, \citenamefont {Sipahigil}, \citenamefont {Nguyen}, \citenamefont
  {Bhaskar}, \citenamefont {Evans}, \citenamefont {Jelezko},\ and\
  \citenamefont {Lukin}}]{Sukachev_2017}%
  \BibitemOpen
  \bibfield  {author} {\bibinfo {author} {\bibfnamefont {D.}~\bibnamefont
  {Sukachev}}, \bibinfo {author} {\bibfnamefont {A.}~\bibnamefont {Sipahigil}},
  \bibinfo {author} {\bibfnamefont {C.}~\bibnamefont {Nguyen}}, \bibinfo
  {author} {\bibfnamefont {M.}~\bibnamefont {Bhaskar}}, \bibinfo {author}
  {\bibfnamefont {R.}~\bibnamefont {Evans}}, \bibinfo {author} {\bibfnamefont
  {F.}~\bibnamefont {Jelezko}}, \ and\ \bibinfo {author} {\bibfnamefont
  {M.}~\bibnamefont {Lukin}},\ }\href@noop {} {\bibfield  {journal} {\bibinfo
  {journal} {Phys. Rev. Lett.}\ }\textbf {\bibinfo {volume} {119}},\ \bibinfo
  {pages} {223602} (\bibinfo {year} {2017})}\BibitemShut {NoStop}%
\bibitem [{\citenamefont {Abobeih}\ \emph {et~al.}(2018)\citenamefont
  {Abobeih}, \citenamefont {Cramer}, \citenamefont {Bakker}, \citenamefont
  {Kalb}, \citenamefont {Markham}, \citenamefont {Twitchen},\ and\
  \citenamefont {Taminiau}}]{Abobeih_2018}%
  \BibitemOpen
  \bibfield  {author} {\bibinfo {author} {\bibfnamefont {M.}~\bibnamefont
  {Abobeih}}, \bibinfo {author} {\bibfnamefont {J.}~\bibnamefont {Cramer}},
  \bibinfo {author} {\bibfnamefont {M.}~\bibnamefont {Bakker}}, \bibinfo
  {author} {\bibfnamefont {N.}~\bibnamefont {Kalb}}, \bibinfo {author}
  {\bibfnamefont {M.}~\bibnamefont {Markham}}, \bibinfo {author} {\bibfnamefont
  {D.}~\bibnamefont {Twitchen}}, \ and\ \bibinfo {author} {\bibfnamefont
  {T.}~\bibnamefont {Taminiau}},\ }\href@noop {} {\bibfield  {journal}
  {\bibinfo  {journal} {Nat. Commun.}\ }\textbf {\bibinfo {volume} {9}},\
  \bibinfo {pages} {2552} (\bibinfo {year} {2018})}\BibitemShut {NoStop}%
\bibitem [{\citenamefont {Jahnke}\ \emph {et~al.}(2012)\citenamefont {Jahnke},
  \citenamefont {Naydenov}, \citenamefont {Teraji}, \citenamefont {Koizumi},
  \citenamefont {Umeda}, \citenamefont {Isoya},\ and\ \citenamefont
  {Jelezko}}]{Jahnke_2012}%
  \BibitemOpen
  \bibfield  {author} {\bibinfo {author} {\bibfnamefont {K.~D.}\ \bibnamefont
  {Jahnke}}, \bibinfo {author} {\bibfnamefont {B.}~\bibnamefont {Naydenov}},
  \bibinfo {author} {\bibfnamefont {T.}~\bibnamefont {Teraji}}, \bibinfo
  {author} {\bibfnamefont {S.}~\bibnamefont {Koizumi}}, \bibinfo {author}
  {\bibfnamefont {T.}~\bibnamefont {Umeda}}, \bibinfo {author} {\bibfnamefont
  {J.}~\bibnamefont {Isoya}}, \ and\ \bibinfo {author} {\bibfnamefont
  {F.}~\bibnamefont {Jelezko}},\ }\href@noop {} {\bibfield  {journal} {\bibinfo
   {journal} {Appl. Phys. Lett.}\ }\textbf {\bibinfo {volume} {101}},\ \bibinfo
  {pages} {021405} (\bibinfo {year} {2012})}\BibitemShut {NoStop}%
\bibitem [{\citenamefont {Nickerson}\ \emph {et~al.}(2014)\citenamefont
  {Nickerson}, \citenamefont {Fitzsimons},\ and\ \citenamefont
  {Benjamin}}]{Nickerson_2014}%
  \BibitemOpen
  \bibfield  {author} {\bibinfo {author} {\bibfnamefont {N.~H.}\ \bibnamefont
  {Nickerson}}, \bibinfo {author} {\bibfnamefont {J.~F.}\ \bibnamefont
  {Fitzsimons}}, \ and\ \bibinfo {author} {\bibfnamefont {S.~C.}\ \bibnamefont
  {Benjamin}},\ }\href@noop {} {\bibfield  {journal} {\bibinfo  {journal}
  {Phys. Rev. X}\ }\textbf {\bibinfo {volume} {4}},\ \bibinfo {pages} {041041}
  (\bibinfo {year} {2014})}\BibitemShut {NoStop}%
\bibitem [{\citenamefont {Choi}\ \emph {et~al.}(2019)\citenamefont {Choi},
  \citenamefont {Pant}, \citenamefont {Guha},\ and\ \citenamefont
  {Englund}}]{Choi_2019}%
  \BibitemOpen
  \bibfield  {author} {\bibinfo {author} {\bibfnamefont {H.}~\bibnamefont
  {Choi}}, \bibinfo {author} {\bibfnamefont {M.}~\bibnamefont {Pant}}, \bibinfo
  {author} {\bibfnamefont {S.}~\bibnamefont {Guha}}, \ and\ \bibinfo {author}
  {\bibfnamefont {D.}~\bibnamefont {Englund}},\ }\href@noop {} {\bibfield
  {journal} {\bibinfo  {journal} {npj Quant. Inf.}\ }\textbf {\bibinfo {volume}
  {5}},\ \bibinfo {pages} {104} (\bibinfo {year} {2019})}\BibitemShut {NoStop}%
\bibitem [{\citenamefont {Quan}\ \emph {et~al.}(2010)\citenamefont {Quan},
  \citenamefont {Deotare},\ and\ \citenamefont {Loncar}}]{Quan_2010}%
  \BibitemOpen
  \bibfield  {author} {\bibinfo {author} {\bibfnamefont {Q.}~\bibnamefont
  {Quan}}, \bibinfo {author} {\bibfnamefont {P.~B.}\ \bibnamefont {Deotare}}, \
  and\ \bibinfo {author} {\bibfnamefont {M.}~\bibnamefont {Loncar}},\
  }\href@noop {} {\bibfield  {journal} {\bibinfo  {journal} {Appl. Phys.
  Lett.}\ }\textbf {\bibinfo {volume} {96}},\ \bibinfo {pages} {203102}
  (\bibinfo {year} {2010})}\BibitemShut {NoStop}%
\bibitem [{\citenamefont {Alajlan}\ \emph {et~al.}(2019)\citenamefont
  {Alajlan}, \citenamefont {Cojocaru},\ and\ \citenamefont
  {Akimov}}]{Alajlan_2019}%
  \BibitemOpen
  \bibfield  {author} {\bibinfo {author} {\bibfnamefont {A.}~\bibnamefont
  {Alajlan}}, \bibinfo {author} {\bibfnamefont {I.}~\bibnamefont {Cojocaru}}, \
  and\ \bibinfo {author} {\bibfnamefont {A.}~\bibnamefont {Akimov}},\
  }\href@noop {} {\bibfield  {journal} {\bibinfo  {journal} {Opt. Mater.
  Express}\ }\textbf {\bibinfo {volume} {9}},\ \bibinfo {pages} {1678}
  (\bibinfo {year} {2019})}\BibitemShut {NoStop}%
\bibitem [{\citenamefont {Vasco}\ \emph {et~al.}(2019)\citenamefont {Vasco},
  \citenamefont {Savona},\ and\ \citenamefont {Gerace}}]{Vasco_2019}%
  \BibitemOpen
  \bibfield  {author} {\bibinfo {author} {\bibfnamefont {J.}~\bibnamefont
  {Vasco}}, \bibinfo {author} {\bibfnamefont {V.}~\bibnamefont {Savona}}, \
  and\ \bibinfo {author} {\bibfnamefont {D.}~\bibnamefont {Gerace}},\
  }\href@noop {} {\bibfield  {journal} {\bibinfo  {journal} {arXiv}\ ,\
  \bibinfo {pages} {arXiv:1910.10647}} (\bibinfo {year} {2019})}\BibitemShut
  {NoStop}%
\bibitem [{\citenamefont {Zhang}\ \emph {et~al.}(2017)\citenamefont {Zhang},
  \citenamefont {Wang}, \citenamefont {Cheng}, \citenamefont {Shams-Ansari},\
  and\ \citenamefont {Loncar}}]{Zhang_2017}%
  \BibitemOpen
  \bibfield  {author} {\bibinfo {author} {\bibfnamefont {M.}~\bibnamefont
  {Zhang}}, \bibinfo {author} {\bibfnamefont {C.}~\bibnamefont {Wang}},
  \bibinfo {author} {\bibfnamefont {R.}~\bibnamefont {Cheng}}, \bibinfo
  {author} {\bibfnamefont {A.}~\bibnamefont {Shams-Ansari}}, \ and\ \bibinfo
  {author} {\bibfnamefont {M.}~\bibnamefont {Loncar}},\ }\href@noop {}
  {\bibfield  {journal} {\bibinfo  {journal} {Optica}\ }\textbf {\bibinfo
  {volume} {4}},\ \bibinfo {pages} {1536–1537} (\bibinfo {year}
  {2017})}\BibitemShut {NoStop}%
\bibitem [{\citenamefont {Desiatov}\ \emph {et~al.}(2019)\citenamefont
  {Desiatov}, \citenamefont {Shams-Ansari}, \citenamefont {Zhang},
  \citenamefont {Wang},\ and\ \citenamefont {Loncar}}]{Desiatov_2019}%
  \BibitemOpen
  \bibfield  {author} {\bibinfo {author} {\bibfnamefont {B.}~\bibnamefont
  {Desiatov}}, \bibinfo {author} {\bibfnamefont {A.}~\bibnamefont
  {Shams-Ansari}}, \bibinfo {author} {\bibfnamefont {M.}~\bibnamefont {Zhang}},
  \bibinfo {author} {\bibfnamefont {C.}~\bibnamefont {Wang}}, \ and\ \bibinfo
  {author} {\bibfnamefont {M.}~\bibnamefont {Loncar}},\ }\href@noop {}
  {\bibfield  {journal} {\bibinfo  {journal} {Optica}\ }\textbf {\bibinfo
  {volume} {6(3)}},\ \bibinfo {pages} {380–384} (\bibinfo {year}
  {2019})}\BibitemShut {NoStop}%
\bibitem [{\citenamefont {Wan}\ \emph {et~al.}(2020)\citenamefont {Wan} \emph
  {et~al.}}]{Wan_2019}%
  \BibitemOpen
  \bibfield  {author} {\bibinfo {author} {\bibfnamefont {N.~H.}\ \bibnamefont
  {Wan}} \emph {et~al.},\ }\href@noop {} {\bibfield  {journal} {\bibinfo
  {journal} {Nature}\ }\textbf {\bibinfo {volume} {583}},\ \bibinfo {pages}
  {226–231} (\bibinfo {year} {2020})}\BibitemShut {NoStop}%
\bibitem [{\citenamefont {Sun}\ \emph {et~al.}(2016)\citenamefont {Sun},
  \citenamefont {Kim}, \citenamefont {Solomon},\ and\ \citenamefont
  {Waks}}]{Sun_2016}%
  \BibitemOpen
  \bibfield  {author} {\bibinfo {author} {\bibfnamefont {S.}~\bibnamefont
  {Sun}}, \bibinfo {author} {\bibfnamefont {H.}~\bibnamefont {Kim}}, \bibinfo
  {author} {\bibfnamefont {G.}~\bibnamefont {Solomon}}, \ and\ \bibinfo
  {author} {\bibfnamefont {E.}~\bibnamefont {Waks}},\ }\href@noop {} {\bibfield
   {journal} {\bibinfo  {journal} {Nature Nanotech}\ }\textbf {\bibinfo
  {volume} {11}},\ \bibinfo {pages} {539–544} (\bibinfo {year}
  {2016})}\BibitemShut {NoStop}%
\bibitem [{\citenamefont {Kindem}\ \emph {et~al.}(2020)\citenamefont {Kindem},
  \citenamefont {Ruskuc}, \citenamefont {Bartholomew}, \citenamefont {Rochman},
  \citenamefont {Huan},\ and\ \citenamefont {Faraon}}]{Kindem_2020}%
  \BibitemOpen
  \bibfield  {author} {\bibinfo {author} {\bibfnamefont {J.}~\bibnamefont
  {Kindem}}, \bibinfo {author} {\bibfnamefont {A.}~\bibnamefont {Ruskuc}},
  \bibinfo {author} {\bibfnamefont {J.}~\bibnamefont {Bartholomew}}, \bibinfo
  {author} {\bibfnamefont {J.}~\bibnamefont {Rochman}}, \bibinfo {author}
  {\bibfnamefont {Y.}~\bibnamefont {Huan}}, \ and\ \bibinfo {author}
  {\bibfnamefont {A.}~\bibnamefont {Faraon}},\ }\href@noop {} {\bibfield
  {journal} {\bibinfo  {journal} {Nature}\ }\textbf {\bibinfo {volume} {580}},\
  \bibinfo {pages} {2} (\bibinfo {year} {2020})}\BibitemShut {NoStop}%
\bibitem [{\citenamefont {Pogorelov}\ \emph {et~al.}(2021)\citenamefont
  {Pogorelov}, \citenamefont {Feldker}, \citenamefont {Marciniak},
  \citenamefont {Jacob}, \citenamefont {Podlesnic}, \citenamefont {Meth},
  \citenamefont {Negnevitsky}, \citenamefont {Stadler}, \citenamefont
  {Lakhmanskiy}, \citenamefont {Blatt}, \citenamefont {Schindler},\ and\
  \citenamefont {Monz}}]{Pogorelov_2021}%
  \BibitemOpen
  \bibfield  {author} {\bibinfo {author} {\bibfnamefont {I.}~\bibnamefont
  {Pogorelov}}, \bibinfo {author} {\bibfnamefont {T.}~\bibnamefont {Feldker}},
  \bibinfo {author} {\bibfnamefont {C.~D.}\ \bibnamefont {Marciniak}}, \bibinfo
  {author} {\bibfnamefont {G.}~\bibnamefont {Jacob}}, \bibinfo {author}
  {\bibfnamefont {V.}~\bibnamefont {Podlesnic}}, \bibinfo {author}
  {\bibfnamefont {M.}~\bibnamefont {Meth}}, \bibinfo {author} {\bibfnamefont
  {V.}~\bibnamefont {Negnevitsky}}, \bibinfo {author} {\bibfnamefont
  {M.}~\bibnamefont {Stadler}}, \bibinfo {author} {\bibfnamefont
  {K.}~\bibnamefont {Lakhmanskiy}}, \bibinfo {author} {\bibfnamefont
  {R.}~\bibnamefont {Blatt}}, \bibinfo {author} {\bibfnamefont
  {P.}~\bibnamefont {Schindler}}, \ and\ \bibinfo {author} {\bibfnamefont
  {T.}~\bibnamefont {Monz}},\ }\href@noop {} {\bibfield  {journal} {\bibinfo
  {journal} {arXiv}\ ,\ \bibinfo {pages} {arXiv:2101.11390}} (\bibinfo {year}
  {2021})}\BibitemShut {NoStop}%
\bibitem [{\citenamefont {Omran}\ \emph {et~al.}(2019)\citenamefont {Omran},
  \citenamefont {Levine}, \citenamefont {Keesling}, \citenamefont {Semeghini},
  \citenamefont {Wang}, \citenamefont {Ebadi}, \citenamefont {Bernien},
  \citenamefont {Zibrov}, \citenamefont {Pichler}, \citenamefont {Choi},
  \citenamefont {Cui}, \citenamefont {Rossignolo}, \citenamefont {Rembold},
  \citenamefont {Montangero}, \citenamefont {Calarco}, \citenamefont {Endres},
  \citenamefont {Greiner}, \citenamefont {Vuletic},\ and\ \citenamefont
  {Lukin}}]{Omran_2019}%
  \BibitemOpen
  \bibfield  {author} {\bibinfo {author} {\bibfnamefont {A.}~\bibnamefont
  {Omran}}, \bibinfo {author} {\bibfnamefont {H.}~\bibnamefont {Levine}},
  \bibinfo {author} {\bibfnamefont {A.}~\bibnamefont {Keesling}}, \bibinfo
  {author} {\bibfnamefont {G.}~\bibnamefont {Semeghini}}, \bibinfo {author}
  {\bibfnamefont {T.~T.}\ \bibnamefont {Wang}}, \bibinfo {author}
  {\bibfnamefont {S.}~\bibnamefont {Ebadi}}, \bibinfo {author} {\bibfnamefont
  {H.}~\bibnamefont {Bernien}}, \bibinfo {author} {\bibfnamefont {A.~S.}\
  \bibnamefont {Zibrov}}, \bibinfo {author} {\bibfnamefont {H.}~\bibnamefont
  {Pichler}}, \bibinfo {author} {\bibfnamefont {S.}~\bibnamefont {Choi}},
  \bibinfo {author} {\bibfnamefont {J.}~\bibnamefont {Cui}}, \bibinfo {author}
  {\bibfnamefont {M.}~\bibnamefont {Rossignolo}}, \bibinfo {author}
  {\bibfnamefont {P.}~\bibnamefont {Rembold}}, \bibinfo {author} {\bibfnamefont
  {S.}~\bibnamefont {Montangero}}, \bibinfo {author} {\bibfnamefont
  {T.}~\bibnamefont {Calarco}}, \bibinfo {author} {\bibfnamefont
  {M.}~\bibnamefont {Endres}}, \bibinfo {author} {\bibfnamefont
  {M.}~\bibnamefont {Greiner}}, \bibinfo {author} {\bibfnamefont
  {V.}~\bibnamefont {Vuletic}}, \ and\ \bibinfo {author} {\bibfnamefont
  {M.~D.}\ \bibnamefont {Lukin}},\ }\href@noop {} {\bibfield  {journal}
  {\bibinfo  {journal} {Science}\ }\textbf {\bibinfo {volume} {365}},\ \bibinfo
  {pages} {570–574} (\bibinfo {year} {2019})}\BibitemShut {NoStop}%
\bibitem [{\citenamefont {Barbarossa}\ \emph {et~al.}(1995)\citenamefont
  {Barbarossa}, \citenamefont {Matteo},\ and\ \citenamefont
  {Armenise}}]{Barbarossa_1995}%
  \BibitemOpen
  \bibfield  {author} {\bibinfo {author} {\bibfnamefont {G.}~\bibnamefont
  {Barbarossa}}, \bibinfo {author} {\bibfnamefont {A.}~\bibnamefont {Matteo}},
  \ and\ \bibinfo {author} {\bibfnamefont {M.}~\bibnamefont {Armenise}},\
  }\href@noop {} {\bibfield  {journal} {\bibinfo  {journal} {J. Lightwave
  Technol.}\ }\textbf {\bibinfo {volume} {13}},\ \bibinfo {pages} {148–157}
  (\bibinfo {year} {1995})}\BibitemShut {NoStop}%
\end{thebibliography}%

\end{document}